\documentclass[journal,comsoc,a4paper]{IEEEtran}

\usepackage[utf8]{inputenc}
\usepackage[T1]{fontenc}

\usepackage{url}
\usepackage[hidelinks]{hyperref}

\usepackage{cite}

\usepackage[dvips]{graphicx}
\graphicspath{{fig/}{pictures/}{fig/out/}}
\DeclareGraphicsExtensions{.png, .jpg, .pdf}

\usepackage[fleqn]{amsmath}
\usepackage{txfonts}
\usepackage{bm}
\usepackage{algorithmic}
\usepackage{array}
\usepackage{commath}

\usepackage[font=footnotesize]{subfig}
\usepackage{caption}
\usepackage{dblfloatfix}
\usepackage{tabularx}
\usepackage{multirow}
\usepackage{xcolor}
\usepackage[export]{adjustbox}
\usepackage[innerleftmargin = 5pt, innerrightmargin = 5pt,innertopmargin = 5pt, innerbottommargin = 5pt]{mdframed}
\usepackage{xcolor, colortbl}
\usepackage{tikz}
\usepackage{collcell}
\usepackage[rightcaption]{sidecap}

\definecolor{shadecolor}{gray}{.95}
\definecolor{green}{RGB}{157,187,97}
\definecolor{red}{RGB}{192,80,70}
\definecolor{lightgray}{RGB}{192,192,192}
\newcolumntype{g}{>{\columncolor{green}\centering\arraybackslash}X}
\newcolumntype{r}{>{\columncolor{red}\centering\arraybackslash}X}
\newcolumntype{a}{>{\columncolor{lightgray}\centering\arraybackslash}c}
\newcolumntype{m}{>{\centering\arraybackslash}X}
\newcolumntype{M}{>{\collectcell\ApplyGradient}c<{\endcollectcell}}
\renewcommand{\arraystretch}{1.5}
\newcommand*{\MinNumber}{-60}
\newcommand*{\MaxNumber}{0}
\newcommand\ifnumber[1]{%
	\begingroup
	\edef\temp{#1}%
	\expandafter\expandafter{\temp}
	{\endgroup\@secondoftwo}
	{\expandafter\ifnumber@i\temp\@nnil}%
}
\def\gobbleminus#1{\ifx-#1\else#1\fi}

\newcommand{\ApplyGradient}[1]{%
		\pgfmathsetmacro{\PercentColor}{100.0*(max(-60,#1)-\MinNumber)/(\MaxNumber-\MinNumber)}
		\hspace{-0.33em}{\adjustbox{ bgcolor=red!\PercentColor!green}{\marginbox{3.8mm 1.5mm}{#1}}}
}
\usepackage{enumerate}

\title{On Spectral Coexistence of CP-OFDM \\and FB-MC Waveforms in 5G Networks}

\author{Quentin~Bodinier,~\IEEEmembership{Student Member,~IEEE,}
        Faouzi~Bader,~\IEEEmembership{Senior~Member,~IEEE,}
        and~Jacques~Palicot,~\IEEEmembership{Member,~IEEE}
\thanks{Quentin Bodinier, Jacques Palicot and Faouzi Bader are with SCEE Research Team at
	CentraleSupélec/IETR, 35576 Cesson-Sévigné, France.}%
\thanks{E-mail:\{quentin.bodinier,
		jacques.palicot,faouzi.bader\}@supelec.fr}
\thanks{This work was partially funded through French National Research Agency (ANR) projects ACCENT5 with grant agreement code: ANR-14-CE28-0026-02 and EPHYL with grant agreement code: ANR-16-CE25-0002-03.}}

\begin{document}
	\pagestyle{empty}
	\onecolumn
	\vspace*{\fill}
	\begin{center}
		\Huge
		\textbf{Notice}\\
		This work has been submitted to the IEEE for possible publication.  Copyright may be transferred without notice, after which this version may no longer be accessible.
	\end{center}
	\vspace*{\fill}
	\twocolumn
	\setcounter{page}{1}
\maketitle

\begin{abstract}
Future 5G networks will serve a variety of applications that will coexist on the same spectral band and geographical area, in an uncoordinated and asynchronous manner. It is widely accepted that using CP-OFDM, the waveform used by most current communication systems, will make it difficult to achieve this paradigm. Especially, CP-OFDM is not adapted for spectral coexistence because of its poor spectral localization. Therefore, it has been widely suggested to use filter bank based multi carrier (FB-MC) waveforms with enhanced spectral localization to replace CP-OFDM. Especially, FB-MC waveforms are expected to facilitate coexistence with legacy CP-OFDM based systems. However, this idea is based on the observation of the PSD of FB-MC waveforms only. In this paper, we demonstrate that this approach is flawed and show what metric should be used to rate interference between FB-MC and CP-OFDM systems. Finally, our results show that using FB-MC waveforms does not facilitate coexistence with CP-OFDM based systems to a high extent.
\end{abstract}


\IEEEpeerreviewmaketitle
\bstctlcite{IEEEexample:BSTcontrol}

\section{Introduction}
\label{sec:intro}
The advent of the 5th Generation of wireless communication systems (5G) is envisioned to bring flexibility to cellular networks. New services as Device-To-Device (D2D) or Machine-To-Machine (M2M) communications are already being progressively deployed. It is expected that the volume of these new communication types, catalyzed by the emergence of the Internet of Things (IoT), will soare in the coming years. The wireless world of tomorrow will therefore be holistically different from the centralized, homogeneous and synchronous cellular networks which abide by the current Long Term Evolution-Advanced (LTE-A) standards \cite{Andrews2014}.

This new paradigm requires the physical layer (PHY) of 5G to be adaptable to various situations, and robust to asynchronous interference coming from neighboring communication devices~\cite{Forschung2013c,Banelli2014}. 

Besides, these new communication types will add a new burden to the radio spectrum, which is already saturated. To answer this challenge, two directions have mainly been explored: 
\begin{enumerate}[(i)]
	\item Exploit new parts of the spectrum at higher frequencies above $6$ GHz.
	\item Exploit parts of the already licensed spectrum that are temporally left free by incumbent users.
\end{enumerate}
Point (i), which is for example related to research on mm-wave communications, is out of the scope of this article. In this study, we rather focus on point (ii). The main challenge to overcome when re-exploiting some parts of the already licensed spectrum lies in the fact that the secondary users should insert their communication without causing harmful interference to the incumbent legacy users. In the context of 5G, a significant part of the reusable spectrum belongs to either LTE-A cellular networks, Wi-Fi, WiMAX or TV bands. In all these cases, inserted secondary users will have to coexist with CP-OFDM based incumbent communications, as it is the waveform used by these technologies. Note that Single Carrier-Frequency Division Multiplexing Access (SC-FDMA) used in the uplink of cellular LTE-A communications is also based on the transmission of Fourier Spread CP-OFDM symbols.

Therefore, future devices deployed in the course of the development of 5G will likely coexist with CP-OFDM based systems. However, it is crucial that newly introduced services do not interfere in a harmful manner with incumbent legacy communications. This issue has been identified as one of the core challenges of cognitive radios by S.~Haykin in \cite{HaykinBrainEmpowered}, where the notion of "interference temperature" was introduced. Currently, to protect the incumbent CP-OFDM users, the common policy states that all users should respect a certain spectrum mask at the transmitter side. In other words, the Power Spectral Density (PSD) of transmitted signals must fit below certain limits specified by the standards. Because of the poor spectral containment of CP-OFDM, it is usually necessary to turn off an important amount of guard subcarriers to fit the specified masks, which incurs an important loss in terms of spectral efficiency. This is a major pitfall which makes coexistence between unsynchronized CP-OFDM systems hardly feasible.

Building on this observation, the research community has investigated the possibility to use new waveforms with improved spectral containment instead of CP-OFDM, so that the secondary transmissions of 5G systems could more easily fit into the spectrum masks specified by incumbent legacy CP-OFDM users. Most new waveform schemes proposed so far fall into the category of filter bank based multi-carrier (FB-MC) waveforms. They all rely on filtering the transmit signal through highly frequency selective filters to reduce out-of-band (OOB) emissions. Therefore, users based on this type of waveform fit more easily the spectral masks set by incumbent users. Based on this observation, an important number of research works have investigated the potential gains obtained by using FB-MC waveforms for coexistence with CP-OFDM based users.

\subsection{Related Work}
The first works related to coexistence of secondary users with CP-OFDM incumbent users date back to the beginning of the century. The cognitive radio concept increasingly gained popularity after Mitola's Ph.D. dissertation \cite{Mitola2000} and the need for a model of interference usable for cognitive radio deployments was soon identified in \cite{HaykinBrainEmpowered}. However, at this time, the research community did not identify clearly the links between this concept and the PHY layer problematics, especially the choice of waveform. It is only when Weiss and Jondral coined the term "spectrum pooling" in \cite{Weiss2004} that they defined a waveform-dependent model to compute interference between systems coexisting on the same spectral band. The model they defined went on to be known as the "PSD-based model" in the literature. 

This model has then been extensively used to compute interference between secondary and incumbent users in cognitive radio scenarios in which both systems would utilize CP-OFDM \cite{Weiss2004, shaat2010computationally}. Because of the high OOB emissions of CP-OFDM, research works based on the PSD-based model showed that CP-OFDM is not well adapted for spectrum sharing between secondary and incumbent users \cite{Mahmoud2009}. This observation was one of the many arguments that fostered the research of a new waveform able to replace CP-OFDM in 5G systems. 

All FB-MC multicarrier waveforms proposed to replace CP-OFDM rely on some filtering to improve the spectral localization of the emitted signals. Some of them perform a per-band filtering, meaning that groups of subcarriers are passed together through a filter. Among this group of subband-filtered waveforms, the two leading proposals are filtered-OFDM (f-OFDM) \cite{Abdoli2015} and universal filtered OFDM (UF-OFDM) originally called universal filtered multi-carrier (UFMC) \cite{Wild2014}. In this paper however, we are interested in subcarrier-filtered waveforms which perform, as their name indicates, a subcarrier-wise filtering on the transmitted signal so that each waveform is highly spectrally localized. Among these waveforms, we consider Filtered Multi-Tone (FMT), offset quadrature amplitude modulated-OFDM (OFDM/OQAM), FBMC with pulse amplitude modulation (FBMC-PAM), generalized frequency division multiplexing (GFDM) and FBMC/circular OQAM(COQAM). \cite{Tonello2005, FMT_Emphatic, Siohan2002, Farhang-Boroujeny2011, Bellanger2015, Fettweis2009, Michailow2014, Lin2014b}

In FMT modulation \cite{Tonello2005, FMT_Emphatic}, a guard band is added between every subcarrier so that they do not overlap. Therefore, FMT suffers from some bandwidth efficiency loss. To increase spectral efficiency, OFDM/OQAM \cite{Siohan2002, Farhang-Boroujeny2011} allows for adjacent subcarriers to overlap. Unlike FMT, real symbols modulated according to a PAM constellation are transmitted on each subcarrier. A $\frac{\pi}{2}$ phase difference is applied to adjacent subcarriers, which provides real-domain orthogonality. However, OFDM/OQAM requires doubling the symbol rate. To avoid that, FBMC-PAM modulation has been recently proposed by Bellanger et al. in \cite{Bellanger2015}. In this scheme, the number of subcarriers is doubled instead of the number of time symbols. A drawback common to all the aforementioned modulations is the transient imposed by their transmit and receive filters. GFDM tackles this problem by application of cyclic pulse shaping \cite{Fettweis2009}. However, this comes at the expense of higher OOB emissions, which is due do the discontinuities induced in the signal by truncation in time with a rectangular window \cite{Michailow2014}. Finally, COQAM consists in a mix between GFDM and OFDM/OQAM \cite{Lin2014b} : OQAM symbols are transmitted through a set of circular filter banks.

Whatever the specific details of the aforementioned waveforms, they all achieve lower OOB emissions than CP-OFDM thanks to their inherent filtering of the transmitted symbols. Because of these advantageous PSD properties, an important amount of research works have investigated the benefits of using these FB-MC waveforms for coexistence with legacy CP-OFDM incumbent users, for example \cite{Noguet2011, shaat2010computationally, skrzypczak2012ofdm, Su2016, Michailow2011}. Interestingly, to rate interference between coexisting systems, many of these studies applied the PSD-based model in heterogeneous scenarios in which the secondary and incumbent systems do not use the same waveform, even though this model was originally designed to analyze scenarios in which both systems use CP-OFDM. 

Aforementioned studies built on the PSD-based model predicted that it would be very efficient to use FB-MC waveforms to coexist with CP-OFDM users. Because of these encouraging results, a number of real-world demonstrations of coexistence between FB-MC waveforms and CP-OFDM have also been undertaken by both industrials and academics \cite{KeysightWaveforms, EurecomExp, EURECOM+4725}. In \cite{KeysightWaveforms}, the error vector magnitude (EVM) of the incumbent CP-OFDM receiver is studied, but the achieved values are not compared for multiple secondary waveforms. Therefore, this experiment does not enable us to rate the potential gains of using FB-MC for coexistence. In \cite{EurecomExp, EURECOM+4725}, the authors measure experimentally the layer 3 goodput of a LTE uplink receiver confronted with interference coming from an asynchronous secondary transmitter. They show that if the latter uses GFDM, it can transmit approximately with 3dB more power than a secondary that would use OFDM. Though this is an interesting gain, it is fairly less significant than what would be expected with the PSD-based model. In \cite{Berg2014743}, the authors study the coexistence of a secondary OFDM/OQAM system with a TV receiver and show that it could transmit with 9dB more power than a CP-OFDM secondary system without destroying the TV signal. Once again, this gap seems interesting but is not in line with results obtained with the PSD-based model for example in \cite{shaat2010computationally} in which it is predicted that FB-MC systems can be assigned almost arbitrary high power without interfering on the primary if just one guard subcarrier is used.

Overall, these experiments rely on high-level and sometimes qualitative metrics which encompass many effects and do not provide reproducible analysis. Moreover, they do not seem to be consistent with the PSD-based approach.
Therefore, this shows that referring to the PSD of the secondary signal to compute interference in heterogeneous scenarios is not sufficient and a novel approach is needed. Moreover, the limitations of the PSD-based model, even in homogeneous scenarios, have been shown by Medjahdi \textit{et al.} in \cite{Medjahdi2010a}.
However, to the best of our knowledge, only few research works have tackled the problem of coexistence between systems with different waveforms in an alternative manner. Most recently, Ahmed \textit{et al.} have analyzed in \cite{Ahmed2016} the coexistence between UF-OFDM and CP-OFDM systems, but did not compare the results obtained when the secondary system uses UF-OFDM with those obtained if it was using CP-OFDM.

\subsection{Contributions}

\footnote{Note to the reader: this article aims at closing a research study on the topic of coexistence between FB-MC and CP-OFDM that we have been leading for one year. Therefore, this article includes some content that has been previously published in international conferences. Namely, in \cite{Bodinier2016ICT}, we showed the limitations of the PSD-based model through simulations. In \cite{Bodinier2016ISWCS}, we explained how some results available in the literature on that matter should be revised based on our observations. Finally, in \cite{Bodinier2016GC}, we derive the analytical expression of interference between OFDM/OQAM and CP-OFDM. The present article therefore compiles, completes and extends these works. Especially, we extend these works which were focused on OFDM/OQAM to any subcarrier filtered FB-MC waveform.}
Based on the above review of the state of the art, the conclusion is clear: even though the research community has been extensively studying coexistence between FB-MC waveforms and CP-OFDM with both simulation and experimental approaches, there is still no accurate analytical model of interference available in the literature to study these scenarios, with the exception of the work in \cite{Ahmed2016} for UF-OFDM. Moreover, it has been noted that precise analysis of coexistence between CP-OFDM and FB-MC waveforms is an open issue that is important for various fields of research and that needs to be addressed \cite{Wang2016Light}. We therefore propose to fill this gap in this article by accurately analyzing the interference between FB-MC waveforms and CP-OFDM in spectral coexistence scenarios. More precisely, in our discussion, we point out that looking at the PSD of the interfering signal is not sufficient because it does not encompass the operations performed by the receiver, as PSD is measured before the input antenna of the receiver that suffers from interference. We recall that interference should be measured after the receiver operations, based on EVM measurement. Based on this approach, our main contributions are two-fold:

\begin{figure}[!t]
	\centering
	\includegraphics[width=\linewidth]{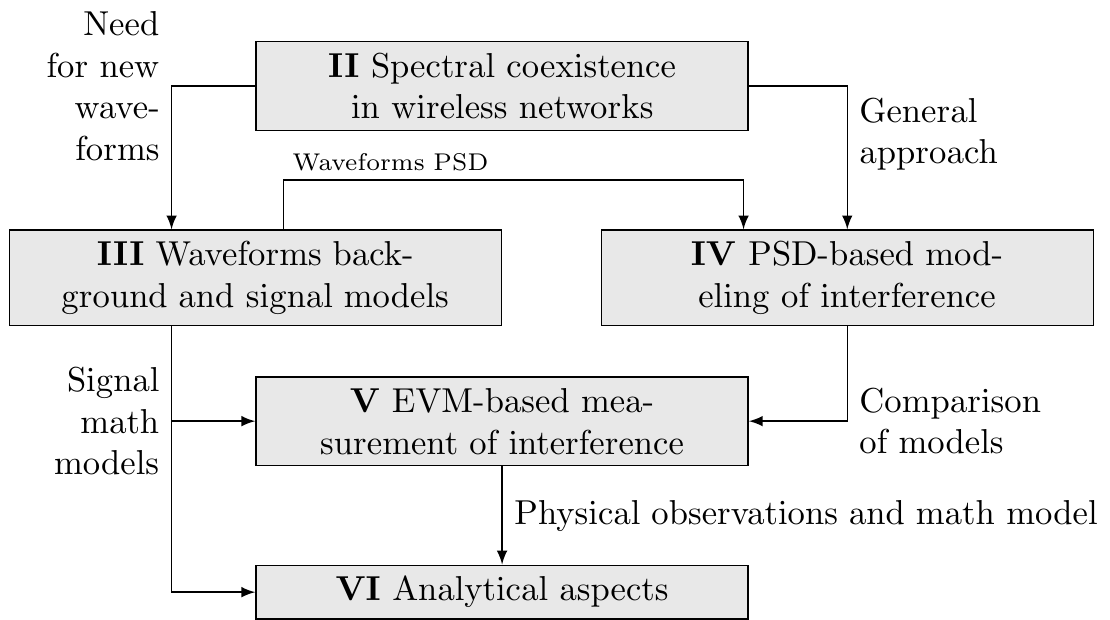}
	\caption{Organization of our analysis.}
	\label{fig:synopsis}
\end{figure}

\begin{enumerate}
	\item We invalidate the PSD-based model and the results based on it for the analysis of scenarios in which systems based on different waveforms coexist.
	\item We provide an analytical framework which explains why FB-MC waveforms do not significantly improve coexistence with CP-OFDM.
\end{enumerate}

This article is organized as described in Fig.~\ref{fig:synopsis}: in section \ref{sec:coexistence}, we present the idea of spectral coexistence in wireless networks, how it is usually tackled in the literature and the system model we propose to study it. Based on this, we explain why new waveforms have been sought, and we present the different waveform schemes we consider in section \ref{sec:waveforms}. In Section \ref{sec:model} we detail the PSD-based model that is commonly used to study those scenarios and show how it applies to the waveforms we consider. In section \ref{sec:interf}, we propose a new measure to rate interference with more accuracy in coexistence scenarios. In section \ref{sec:analysis}, we derive analytical expressions of interference according to the newly proposed model. 
Finally, section \ref{sec:ccl} concludes this article.

\textbf{Notations :} scalars are noted $x$, vectors are bold-faced as $\mathbf{x}$ and sets are written with calligraphic letters $\mathcal{X}$. $t$ represents continuous time, whereas $m$ and $n$ index respectively the subcarriers and the time slots. $\{.\}^*$ represent the complex conjugate, and $E_x\{y\}$ represents the expected value of  $y$ with respect to random variable $x$. Finally, $\overline{\mathbf{x}}$ represents the average value of vector $\mathbf{x}$.

\section{Spectral coexistence in wireless networks}
\label{sec:coexistence}

\subsection{Spectral coexistence : definition}
Coexistence in wireless networks is a notion that can take many meanings and forms; therefore, we feel that it is important to introduce our discussion by stressing out what \textbf{we} mean by spectral coexistence in wireless networks. For example, there has recently been a soaring interest for coexistence between Wi-Fi and LTE-M. This coexistence is enabled at the medium access control (MAC) layer through listen before talk procedures. This is \textbf{not} the type of coexistence that we are considering in this article.

Rather, the type of coexistence we cover in this work is similar to what was called "spectrum pooling" in the past \cite{Weiss2004}, in which two systems coexist in the same spectral band, and each of them is assigned a specific part of the said band. An important assumption in our work is that the two coexisting systems operate independently, with no synchronization and/or cooperation between them. Besides, note that our study is topology-agnostic. The only assumption we make is that a secondary system transmits on spectral resources adjacent to a CP-OFDM based incumbent, and we then explore the consequences of the secondary utilizing a FB-MC waveform on the coexistence capabilities of the considered users.

\begin{figure*}[]
	\centering
	\captionsetup{justification=justified}
	\subfloat[]{\includegraphics[width=0.33\linewidth]{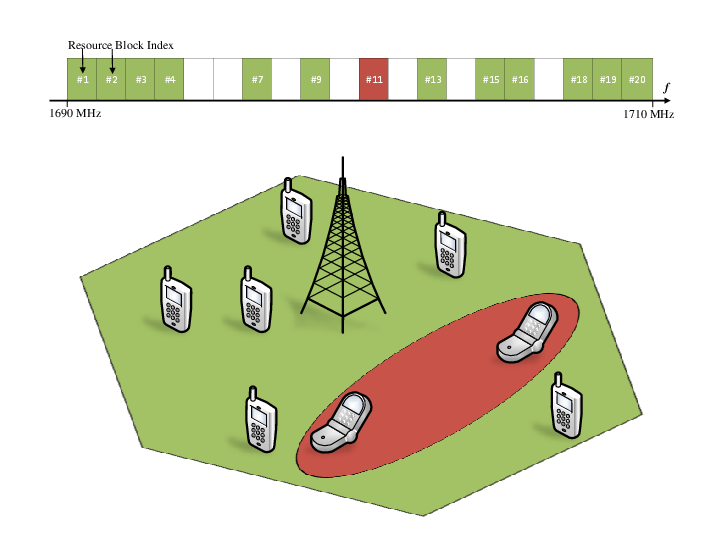}\label{fig:auton_d2d}}
	\subfloat[]{\includegraphics[width=0.33\linewidth]{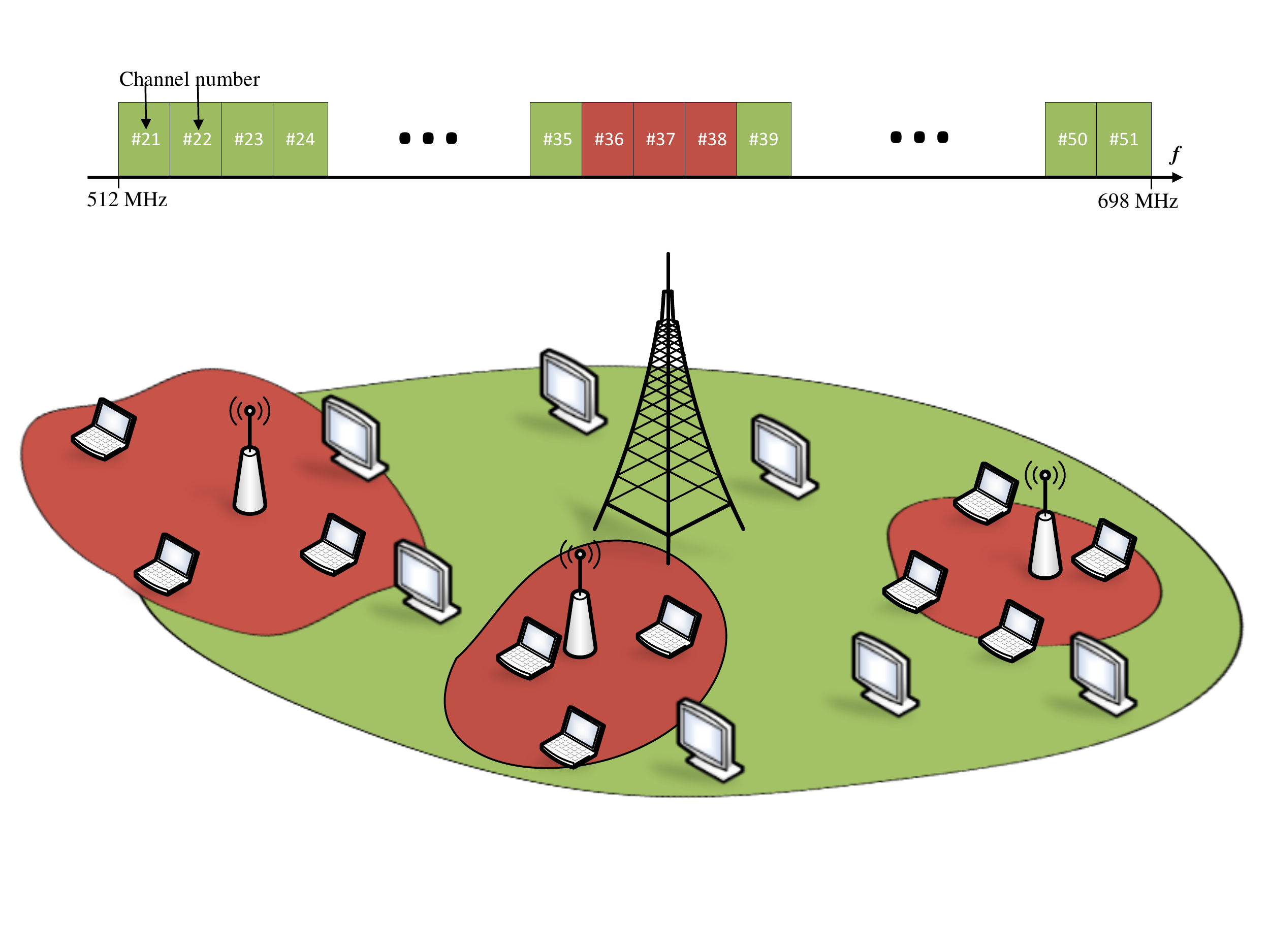}\label{fig:wlan_tvws}}
	\subfloat[]{\includegraphics[width=0.33\linewidth]{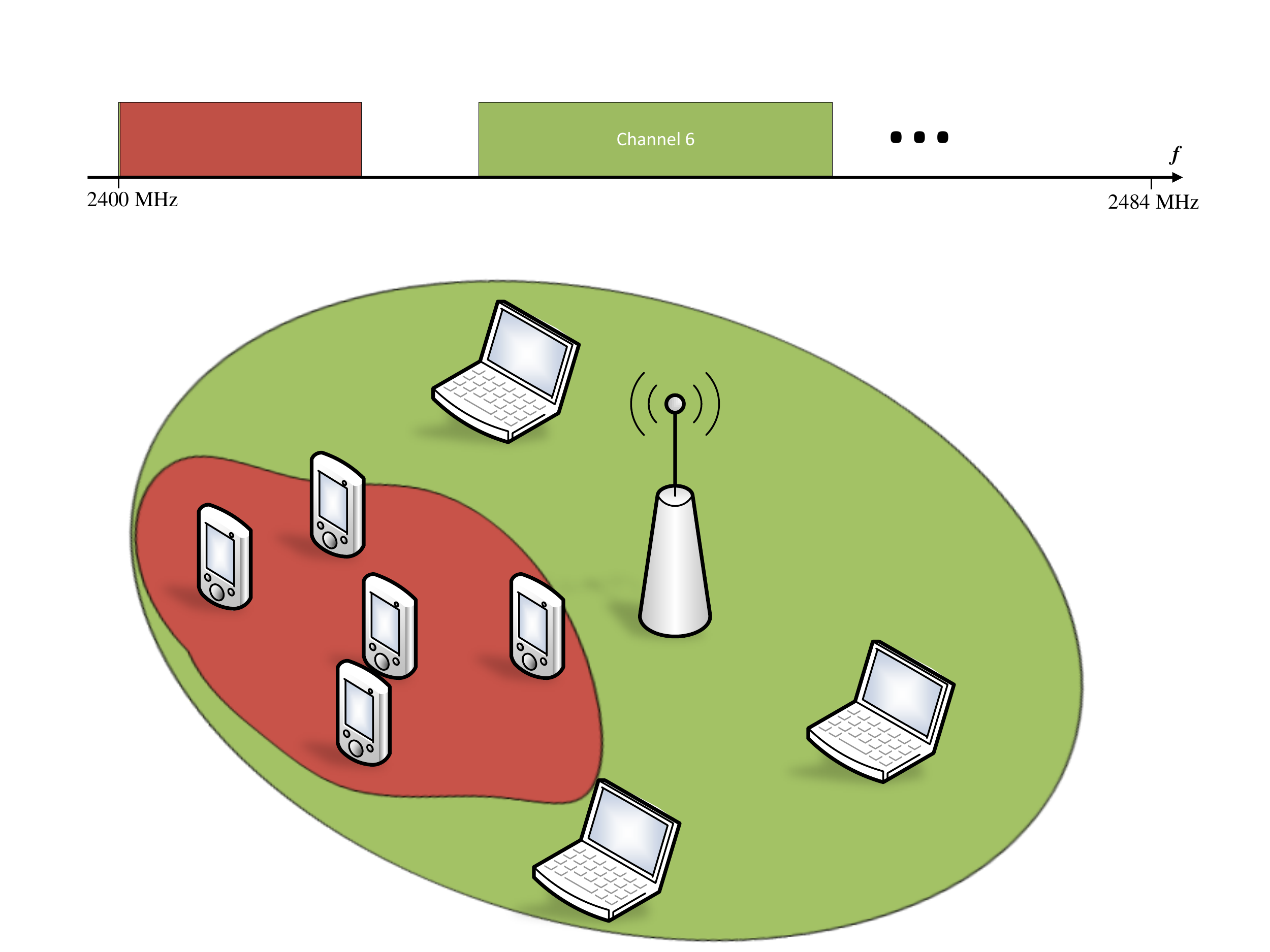}\label{fig:mesh_wifi}}
	\caption{Examples of scenarios where the analysis of this article is relevant:\newline(a) Insertion of an autonomous D2D link in a free resource block in the uplink band of a LTE cellular network \newline(b) Establishment of WLAN cells in TV white spaces\newline(c) Coexistence of ad-hoc mesh network with a Wi-Fi access point}
	\label{fig:deployment_scenarios}
\end{figure*}

Note that such a coexistence scenario can occur in multiple use cases. A non-comprehensive list of examples is represented in Fig.~\ref{fig:deployment_scenarios}. For instance, Fig~\ref{fig:auton_d2d} encompasses a situation in which a cellular LTE network is deployed and one of the free resource blocks (RBs) of its uplink band is reused by an autonomous D2D link between two devices. This kind of policy is of prime interest as it enables the network to use resources which would otherwise be left vacant. However, in that particular case, the D2D link could cause harmful interference to the LTE base station. Such scenarios have been investigated in \cite{BodinierICC2016, Xing2014}. In Fig~\ref{fig:wlan_tvws}, a situation in which some wireless local area networks (WLAN) are deployed in TV white spaces (TVWS) is represented. In that case, WLAN users could harm TV reception of primary users. This type of coexistence scenario has been widely investigated since the FCC took the decision to allow secondary communications to take place in TV white spaces \cite{FCConWhiteSpaces, Noguet2011, Berg2014743}. In Fig.~\ref{fig:mesh_wifi}, we give a last example of a mesh ad-hoc network utilizing the 2.4 GHz industrial, scientific and medical (ISM) band, thus causing harmful interference to a neighboring Wi-Fi access point. In all of these examples, application and network topology vary, but one thing remains: a secondary system coexists with a CP-OFDM based incumbent one on the same frequency band.

\subsection{Limitation of current coexistence studies}
\label{sec:spectrum_masks}

\begin{figure}
	\includegraphics[width=\linewidth]{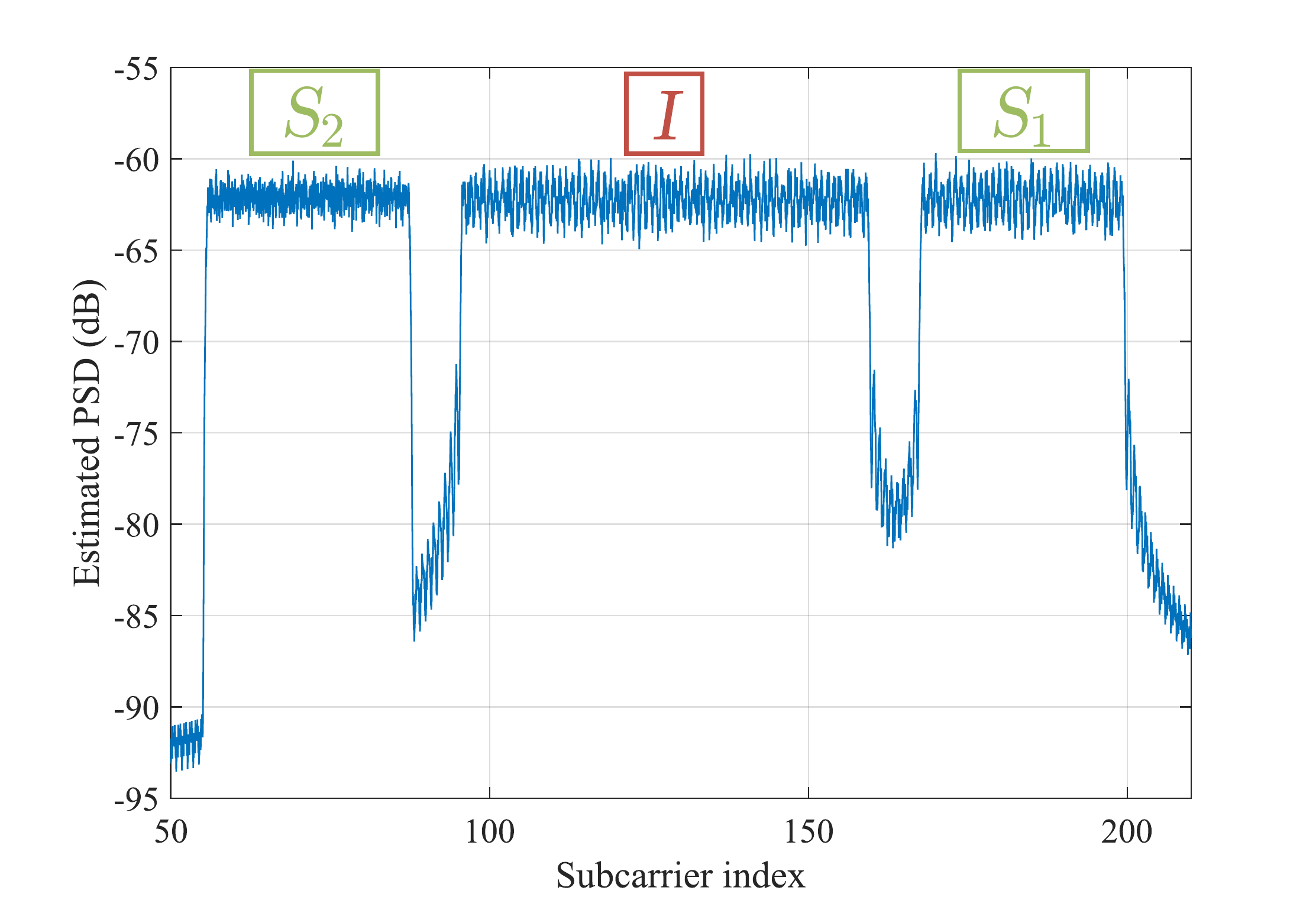}
	\caption{Typical example of a coexistence study in the literature: one incumbent system $\textsl{I}$ coexists with two secondary systems, $\textsl{S}_1$ on its right and $\textsl{S}_2$ on its left. Because $\textsl{S}_2$ has better spectral localization, it is expected that it will not hinder the reception of system $\textsl{I}$ as much as  $\textsl{S}_1$ will.
	}
	\label{fig:classical_experiment}
\end{figure}

In this article, we do not claim to adress a new problem: a number of studies analyzing coexistence between various waveforms in heterogeneous scenarios have been led in the literature \cite{shaat2010computationally, Ahmed2016, Su2016, Noguet2011, skrzypczak2012ofdm}. Rather, we claim that the way these analysis have been led so far is insufficient, and propose a new approach. Indeed, most available studies simply rely on the observation of the spectrum to rate the coexistence capabilities of various systems, in a way that is illustrated in Fig.~\ref{fig:classical_experiment}.

Based on the observation of the spectrum only, most studies directly conclude that waveforms with improved spectral localization are best suited to coexist with CP-OFDM based incumbent systems. These works rely on the PSD-based model that we will introduce in Section \ref{sec:model}. However, with this approach, the gains are difficult to quantize precisely, and more accurate models are needed.
In the following, we therefore introduce a simple model that we will use in the following of this article to study the coexistence between one FB-MC secondary and one incumbent CP-OFDM system.

\subsection{System model}
\label{sec:sys_model}
\begin{figure}[t]
	\centering
	\includegraphics[width=\linewidth]{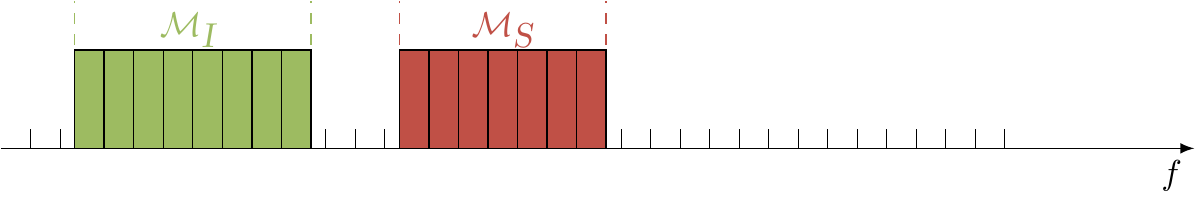}
	\caption{Spectral representation of the considered scenario. The incumbent and secondary systems coexist in the same spectral band, and each one is assigned a different subset of subcarriers. 
}
\label{fig:spec_view}
\end{figure}

\begin{figure}[t]
	\centering
	\includegraphics[width=\linewidth]{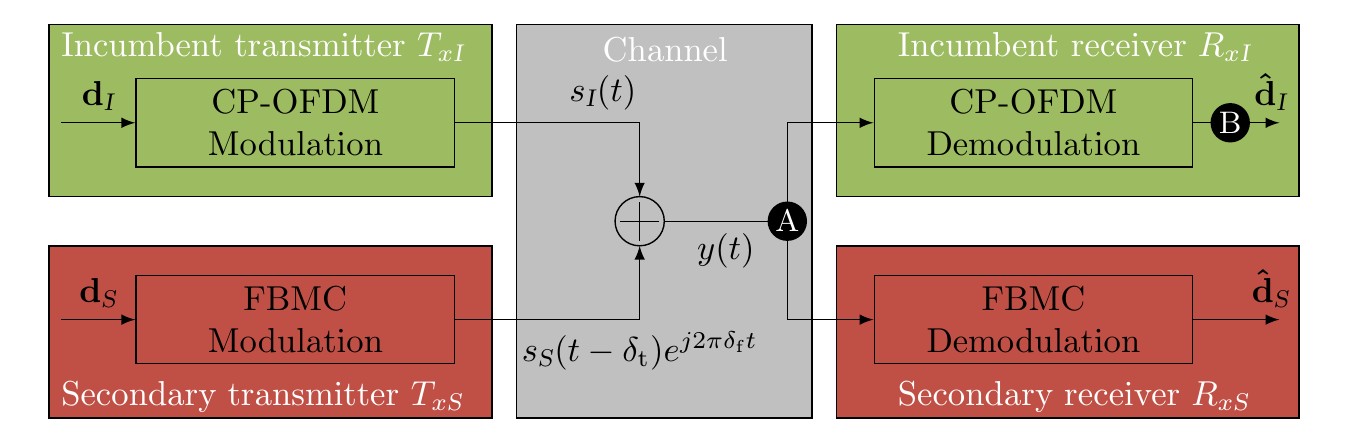}
	\caption{Simple system model to tackle the coexistence between CP-OFDM based incumbent and FB-MC based secondary systems. The channel is modeled as a simple Dirac impulse to focus on inter-system interference caused by the coexistence between the two heterogeneous systems. Timing and frequency misalignment factors are introduced to reflect the fact that there is no inter-system synchronization. Whereas most existing studies rate interference through PSD at point $\vcenter{\hbox{\protect\includegraphics[]{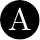}}}$, we propose to analyze it through EVM at point $\vcenter{\hbox{\protect\includegraphics[]{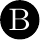}}}$.}
	\label{fig:sys_model}
\end{figure}

Let us lay out a simple system model consisting of a secondary system, \textsl{S}, and an incumbent system \textsl{I}, that coexist in a spectral band and both use a certain set of subcarriers, respectively $\mathcal{M}_\textsl{S}$ and $\mathcal{M}_\textsl{I}$, as is shown in Fig.~\ref{fig:spec_view}. The incumbent system \textsl{I} is composed of a pair of CP-OFDM users, whereas the users forming the secondary system \textsl{S} are based on an alternative FB-MC waveform, as depicted in Fig.~\ref{fig:sys_model}. We show that in each system, the vector $\mathbf{d}$ containing the symbols to be transmitted is fed to the waveform modulator. Then the signals transmitted by both systems are added in the wireless channel and received by both systems which demodulate the received signal and detect the vector of estimated symbols $\hat{\mathbf{d}}$.

Therefore, each system may interfere on the other in function of the properties of the transmitted signals and the demodulation operations that are performed at the receive end.
To show how this model can be used to tackle the various example scenarios presented in Fig.~\ref{fig:deployment_scenarios}, we summarize in Table~\ref{tab:scenarios_match} the roles played by each type of user in each scenario, according to our simplified model. 


\begin{table}[t]
	\caption{Application of proposed model to example scenarios}
	\label{tab:scenarios_match}
	\centering
	\begin{tabularx}{0.9\linewidth}{|m||g|g|r|r|}
		\hline
		\hline
		\textbf{Scenario} & \multicolumn{2}{>{\columncolor{green}}c|}{\textbf{Incumbent}} & \multicolumn{2}{>{\columncolor{red}}c|}{\textbf{Secondary}}\\
		& $T_{x\textsl{I}}$ & $R_{x\textsl{I}}$ & $T_{x\textsl{S}}$ & $R_{x\textsl{S}}$\\
		\hline
		(a) & Cellular user & Cellular base station & D2D transmitter & D2D receiver\\
		\hline
		(b) & TV broadcast antenna & TV receiver &  WLAN access point & WLAN user \\
		\hline
		(c) & Wi-Fi access point & Wi-Fi user & Ad-hoc user & Ad-hoc user \\
		\hline\hline
	\end{tabularx}
\end{table}

Two main specificities of the system model presented in Fig.~\ref{fig:sys_model} must be pointed out. In the first place, we model the channel very simply. We consider neither multipath effects nor additive white Gaussian noise (AWGN). This is done deliberately to put the focus of this work on inter-system interference.
Besides, as previously mentioned in introduction, a core part of our analysis lies in the fact that the secondary and the incumbent users coexist without pursuing any collaboration and/or synchronization between them. Therefore, it is most likely that both systems will disagree on the time and frequency basis. 

To encompass this effect, we introduce two parameters :
\begin{enumerate}[i.]
	\item \textbf{Time misalignment} $\delta_\text{t}$ : as the incumbent and secondary systems do not synchronize their transmission in time, we assume that the secondary system starts its transmission with a certain random delay $\delta_\text{t}$ with respect to the incumbent system.
	\item \textbf{Frequency misalignment} $\delta_\text{f}$ : local oscillators (LOs) of mobile terminals have a typical accuracy of $\pm 1$ ppm with respect to their nominal frequency \cite{CompSyncConcepts}. For example, at a carrier frequency of $2$ GHz, this can yield a misalignment between users of around $10^{4}$ Hz, which can become significant as it is close to the LTE subcarrier width of $15$~kHz. Therefore, as the incumbent and secondary systems do not cooperate, we assume that the secondary system misaligns its carrier frequency of a factor $\delta_\text{f}$ with respect to the incumbent system. However, note that we consider that the transmitter and receiver in each separate system achieve perfect frequency synchronization.
\end{enumerate}

Taking into account these two factors, the signal received at the input antenna of the incumbent CP-OFDM receiver is, as shown in Fig.~\ref{fig:sys_model},
\begin{equation}
y(t) = s_\textsl{I}(t)+s_\textsl{S}(t-\delta_{\text{t}})e^{j2\pi\delta_\text{f}(t-\delta_{\text{t}})}.
\label{eq:sum_in_channel}
\end{equation}
In order to build up on this system model, we present in the following section the signal models of the CP-OFDM and FB-MC waveforms that the incumbent and secondary systems can use.

\section{Waveforms background and signal models}
\label{sec:waveforms}
\subsection{Generic multicarrier waveform model}
We consider a multi-carrier system with $M$ subcarriers, time spacing between symbols $\Delta \text{T}$ and subcarrier spacing $\Delta\text{F}$. Then, the time-domain continuous baseband signal that is transmitted by this system is written as 
\begin{equation}
s(t) = \sum\limits_{m=0}^{M-1}\underset{s_m(t)}{\underbrace{\sum\limits_{n\in\mathbb{Z}}\mathbf{d}_m[n]f_{n,\text{T}}(t-n\Delta\text{T})e^{j2\pi m \Delta\text{F}(t-n\Delta\text{T})}}}, \forall t \in \mathbb{R},
\label{eq:signal_sum_subcarriers}
\end{equation}
where $\mathbf{d}_m$ is the vector of symbols that are transmitted on the $m$th subcarrier, and $f_{n,\text{T}}$ is the transmit filter used to transmit the $n$th symbol on each subcarrier by the multi-carrier system.

In our work, we assume perfect synchronization in both time and frequency between the transmitter and the receiver of the same system. Under this assumption, at the receiver side, the signal $y(t)$ is demodulated and the vector of estimated symbols on each subcarrier is expressed as
\begin{equation}
\hat{\mathbf{d}}_m[n] = \int\limits_{-\infty}^{+\infty}f_{n,\text{R}}(t-n\Delta\text{T})e^{-j2\pi m\Delta\text{F}(t-n\Delta\text{T})}y(t)\dif t,
\label{eq:general_demod}
\end{equation}
where $f_{n,\text{R}}$ is the receive filter used on each subcarrier to demodulate the $n$th incoming symbol. Based on this generic model, we can describe any waveform, either CP-OFDM or FB-MC, by simply defining the parameters $f_{n,\text{T}}$, $f_{n,\text{R}}$, $\Delta \text{T}$  and $\Delta\text{F}$.
In the following, we configure this generic model for the waveforms of interest to this study.\footnote{Due to space limitations, we can only merely scratch the surface of each presented waveform. We therefore focus on the parameters that are of interest to us in this work and refer the reader to the provided references for further details.}

\subsection{CP-OFDM}
\label{sec:CPOFDM_param}
In CP-OFDM systems, data is drawn from a complex constellation, for example a quadrature amplitude modulation (QAM). Then, incoming data is directly passed through an inverse fast Fourier transform processing block (IFFT). This operation is equivalent to filtering symbols on each subcarrier by a rectangular window in the time domain. Moreover, a CP of duration $T_\text{CP}$ is added to each symbol of duration $T$ before transmission. At the receiver side, the CP is removed and the resulting signal is passed through an FFT block. Therefore, CP-OFDM systems are defined by the following set of parameters:
\begin{itemize}
	\item $\Delta\text{T} = T+T_\text{CP}$
	\item $\Delta\text{F} = \frac{1}{T}$
	\item $\forall n \in \mathbb{Z}, f_{n,\text{T}}(t) = f_\text{T}(t) = \begin{cases}
	\frac{1}{\sqrt{T}}, & t \in [-T_\text{CP}, T]\\
	0, & \text{elsewhere}
	\end{cases}$
	\item $\forall n \in \mathbb{Z}, f_{n,\text{R}}(t) = f_\text{R}(t) = \begin{cases}
	\frac{1}{\sqrt{T}}, & t \in [0, T]\\
	0, & \text{elsewhere}
	\end{cases}$
\end{itemize}

\subsection{Linear convolution based filter banks}
\label{sec:linear_FBMC}
Linear FB-MC systems are configured in a way that is very close to CP-OFDM systems. The main difference between both systems lies in the fact that linear FB-MC systems use a selective prototype filter $g$ of length $T_g$ on each subcarrier to improve their spectral localization compared to CP-OFDM. Note that these systems usually use real symmetric filters so that $g^*(-t) = g(t) \forall t \in \mathbb{R}$.
Therefore, every linear convolution based FB-MC system has the following set of transceive filter:
\begin{itemize}
		\item $\forall n \in \mathbb{Z}, f_{n,\text{T}}(t) = f_\text{T}(t) = \begin{cases}
		g(t), & t \in \left[-\frac{T_g}{2}, \frac{T_g}{2}\right]\\
		0, & \text{elsewhere}
		\end{cases}$
		\item $\forall n \in \mathbb{Z}, f_{n,\text{R}}(t) = f_\text{R}(t) = \begin{cases}
		g(t), & t \in \left[-\frac{T_g}{2}, \frac{T_g}{2}\right]\\
		0, & \text{elsewhere}
		\end{cases}$
\end{itemize}
Based on this structure, linear FB-MC waveforms can come in different flavours according to the  values taken by the input symbols $\mathbf{d}_m[n]$ and the parameters $\Delta\text{T}$ and $\Delta\text{F}$. We specify this in the following.
text
\subsubsection{FMT} \cite{Tonello2005, FMT_Emphatic}
The main drawback of filtering subcarriers with a selective filter is that it destroys the orthogonality between subcarriers and therefore creates harmful inter-carrier interference. Indeed, the Balian-Low Theorem (BLT) states clearly that a multicarrier system cannot have at the same time, ~\cite{Balian1981, Low1985}:
\begin{enumerate}[i.]
	\item good time-frequency localization
	\item Gabor density equal to  $\frac{1}{\Delta\text{T}\Delta\text{F}}$
	\item complex orthogonality between adjacent symbols in time and frequency
\end{enumerate}
In FMT, the harmful inter-carrier interference introduced by prototype filter $g$ is countered by introducing a guard band of width $W_\text{GB}$ between every subcarrier. Therefore, FMT systems use the following set of parameters : \\
$\Delta\text{T} = T, \ \Delta\text{F} = \frac{1}{T}+W_\text{GB}$
\\
Therefore, we see clearly that the Gabor density of FMT systems is lower than 1 which degrades their spectral efficiency.
\subsubsection{OFDM/OQAM}  \cite{Siohan2002, Farhang-Boroujeny2011}
The main problem of FMT being the spectral efficiency loss caused by insertion of a guard subcarrier, the goal of OFDM/OQAM is to remove the latter. However, without these guard bands, as stated by the BLT, complex orthogonality between symbols cannot be achieved. To solve this problem, OFDM/OQAM systems transmit real symbols drawn from a pulse amplitude modulation (PAM). A $\frac{\pi}{2}$ phase difference is then added between subsequent symbols and they are transmitted at twice the symbol rate $\frac{2}{T}$. As a result of this, orthogonality in the real domain is achieved. Therefore, OFDM/OQAM is based on the following set of parameters:\\
$\Delta\text{T} = \frac{T}{2}, \ \Delta\text{F} = \frac{1}{T}$

\subsubsection{FBMC-PAM} \cite{Bellanger2015}
One of the problems of OFDM/OQAM lies in the doubling of the symbol rate. A recent alternative, called FBMC-PAM or Lapped FBMC consists in doubling the number of subcarriers and multiplying each symbol by a certain phase factor $\phi_{m,n} = (n-\frac{1}{2}+\frac{M}{2})(m-\frac{1}{2})$. As a result, this modulation also achieves real orthogonality and has the following parameters: \\
 $\Delta\text{T} = T, \ \Delta\text{F} = \frac{1}{2T}$

\subsection{Circular convolution based filter banks}

\begin{figure}
	\centering
	\includegraphics[trim={2.5cm 0 2.5cm 0},clip,width=\linewidth]{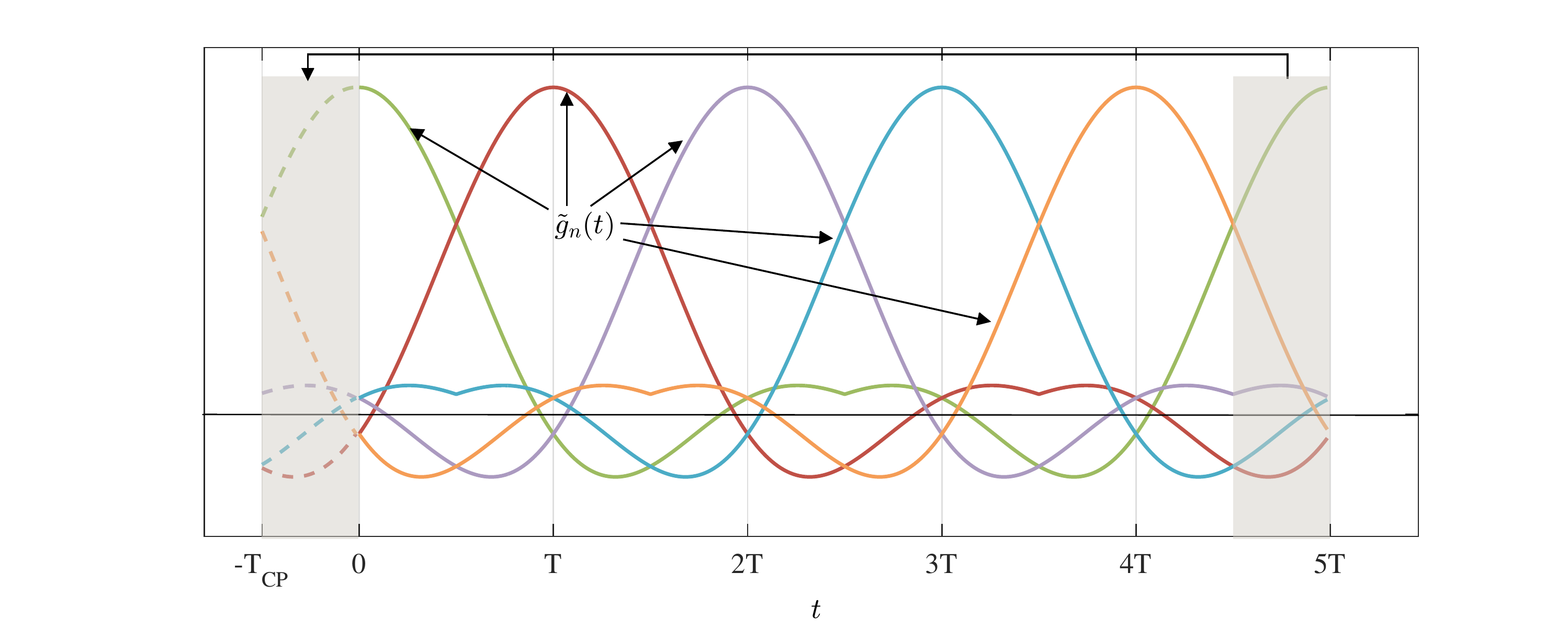}
	\caption{GFDM prototype filters $\tilde{g}_n(t)$ obtained from an original RRC prototype filter $g$ with roll-off 0.25, $N_\text{b} = 5$ and  $T_\text{CP} = \frac{T}{4}$.}
	\label{fig:lineartocirc}
\end{figure}
As explained in introduction, one of the main drawbacks of linear convolution based filter banks is that they incur a delay in the transmission because of the linear convolution with prototype filter $g$. Therefore, following an idea first coined as "tail biting" in \cite{Fettweis2009}, circular convolution based filter banks are based on convolution with circularly shifted versions of the prototype filter $g$. Based on this idea, two main waveforms have been proposed.

\subsubsection{GFDM \cite{Fettweis2009,Michailow2014}} 
In GFDM, complex symbols are modulated per block of $N_\text{b}$ and a CP of length $T_\text{CP}$ is added between every subsequent block. Therefore, from an initial prototype filter $g$, GFDM systems define a set of $N_\text{b}$ filters defined as, $\forall n\in[0, N_\text{b}-1],$
\begin{equation}
\tilde{g}_n(t) = \begin{cases}
g(t-nT \mod N_\text{b}T), & t \in [-T_\text{CP}, N_\text{b}T] \\
0 & \text{elsewhere.}
\end{cases}
\end{equation}
An example of how this applies is given in Fig.~\ref{fig:lineartocirc} where we plot the 5 prototype filters defined from an initial prototype filter $g(t)$ in the case where $g$ is a root-raised cosine (RRC) filter with roll-off 0.25, $N_\text{b} = 5$ and $T_\text{CP} = \frac{T}{4}$.

Based on this set of prototype filters, GFDM systems utilize the following set of transceive filters and parameters:
\begin{itemize}
	\item $\forall n \in \mathbb{Z}, f_{n,\text{T}}(t) = \begin{cases}
	\tilde{g}_{n \mod N_\text{b}}(t), & t \in \left[-T_\text{CP}, N_\text{b}T\right]\\
	0, & \text{elsewhere}
	\end{cases}$
	\item $\forall n \in \mathbb{Z}, f_{n,\text{R}}(t) = f_\text{R}(t) = \begin{cases}
	\tilde{g}_{n \mod N_\text{b}}, & t \in \left[-T_\text{CP},  N_\text{b}T\right]\\
	0, & \text{elsewhere}
	\end{cases}$
	\item $\Delta\text{T} = N_\text{b}T+T_\text{CP},\ \Delta\text{F} = \frac{1}{T}$
\end{itemize}

\subsubsection{COQAM \cite{Lin2014b}}
GFDM suffers from the same limitations imposed by the BLT as linear convolution based FB-MC. Therefore, GFDM based systems need to compensate for inter-carrier interference through specific receiver schemes. Another possibility is to adapt the OQAM modulation to GFDM, a proposal called COQAM. This system is essentially similar to GFDM, the only difference lying in the fact that, as it is based on OQAM, real symbols are transmitted at half the symbol rate. Therefore, according to our model, $2N_\text{b}$ prototype filters are defined from the original prototype filter $g$ as  $\forall n\in[0, 2N_\text{b}-1],$
\begin{equation}
\tilde{g}_n(t) = \begin{cases}
g(t-n\frac{T}{2} \mod N_\text{b}T), & t \in [-T_\text{CP}, N_\text{b}T] \\
0 & \text{elsewhere.}
\end{cases}
\end{equation}
The rest of the parameters are defined exactly as for GFDM.

Note that, because of the circular convolution, GFDM and COQAM have poorer spectral containment than linear convolution based FB-MC waveforms. To solve this problem, a number of research works have proposed to add windowing and/or filtering on top of GFDM or COQAM modulations \cite{Lin2014b, Michailow2014}. We deliberately ignore these solutions in our analysis, as we are interested in studying how the intrinsic filtering properties of the presented waveforms may improve coexistence with CP-OFDM based users.

\subsection{Summary and PSD comparison}
\begin{table}
	\caption{Parameters used by described waveforms}
	\label{tab:waveforms_par}
	\begin{tabularx}{\linewidth}{|c||m|m|m|m|m|m|}
		\hline
		\hline
		\cellcolor{lightgray} & CP-OFDM & FMT & OFDM/\newline OQAM & FBMC-PAM & GFDM & COQAM  \\
		\hline\hline
		 $\Delta \text{F}$ & \multicolumn{3}{c|}{1} & $\frac{1}{2}$ & \multicolumn{2}{c|}{1} \\
		 \hline
		 $T$ &$\frac{1}{\Delta \text{F}}$ & $\frac{1}{\Delta \text{F}-W_\text{GB}}$ & \multicolumn{4}{c|}{$\frac{1}{\Delta \text{F}}$}\\
		 \hline
		 $\Delta \text{T}$ & $T+T_\text{CP}$ & $T$ & $\frac{T}{2}$ & $T$ & \multicolumn{2}{c|}{$N_\text{b}T+T_\text{CP}$} \\
		 \hline
		 $g$ & \cellcolor{lightgray} & RRC, roll-off $1$ & Phydyas \cite{Bellanger} & Sine filter \cite{Bellanger2015} & \multicolumn{2}{c|}{RRC, roll-off $1$} \\
		 \hline
		 $T_g$ & \cellcolor{lightgray} & $6T$ &  $4T$ & $2T$ & \multicolumn{2}{c|}{$5T$} \\
		 \hline
		 $N_\text{b}$ &\multicolumn{4}{a|}{ } &  \multicolumn{2}{c|}{5} \\
		 \hline
		 	 $T_\text{CP}$ & $\frac{T}{8}$ & \multicolumn{3}{a|}{ } & \multicolumn{2}{c|}{$\frac{T}{8}$} \\
		  \hline
		  $W_\text{GB}$ & \cellcolor{lightgray} & $\frac{\Delta\text{F}}{9}$ & \multicolumn{4}{a|}{ }\\
		  \hline
		\hline
	\end{tabularx}
\end{table}

Based on the above presentation of CP-OFDM and FB-MC waveforms, we summarize in Table~\ref{tab:waveforms_par} the set of parameters and prototype filters used by each waveform. The values set in this table will be used in the remaining of this study. In order to have a fair point of comparison, we fix $\Delta \text{F} = 1$ for reference for each waveform, with the exception of FBMC-PAM which uses subcarriers that are only half as wide as those of other systems.

\begin{figure}
	\includegraphics[trim={1cm 0 1cm 0},clip,width=\linewidth]{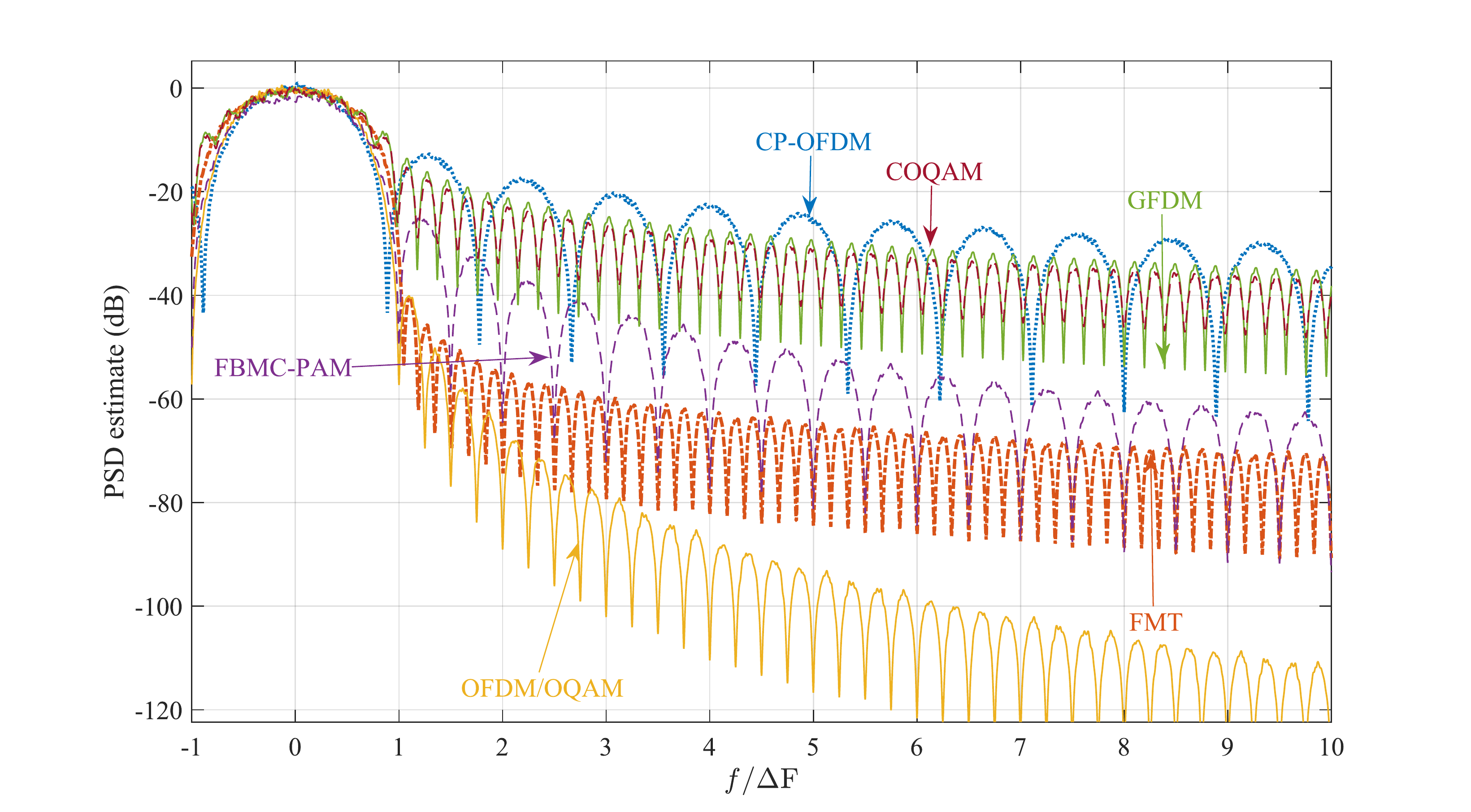}
	\caption{Welch estimate of the PSD of studied waveforms with a Hanning Window of length $\frac{100}{\Delta\text{F}}$ and a frequential resolution of $500$ points per subcarrier.}
	\label{fig:PSD_waveforms}
\end{figure}

In Fig.~\ref{fig:PSD_waveforms}, we present the PSD of one subcarrier of the aforementioned waveforms with the chosen parameters. Note that, to achieve a fair comparison, we considered that not one, but two adjacent subcarriers of the FBMC-PAM system were active. Moreover, it is important to recall that the presented PSD curves correspond to theoretical values that do not take into account hardware impairments and nonlinearities. In real systems, the sidelobes of each waveform would be increased \cite{Renfors2016f}. It appears clearly that circular convolution based FB-MC systems do not achieve satisfying spectral containment, opposed to linear convolution based FB-MC systems whose sidelobes decrease much more rapidly. Following the idea presented in Section \ref{sec:spectrum_masks}, it is expected that the waveforms with better spectral containment will be better suited for coexistence with CP-OFDM based legacy systems. More specifically, at this stage, we can give a tentative ranking of the studied waveforms based on their ability to coexist with CP-OFDM based users. From the sole observation of the PSD curves, OFDM/OQAM appears as a clear winner, followed by FMT and FBMC-PAM in second and third ranking respectively, whereas GFDM, COQAM and CP-OFDM seem to be almost equally bad. This can be quantified through the use of the PSD-based model, which we present in the following.

\section{PSD-based modeling of interference}
\label{sec:model}

\subsection{Model definition}
The PSD-based model consists in computing the leakage caused by users onto each other by integrating the PSD of the interfering signal on the band that suffers from the interference. Therefore, in the scenario that we described in Fig.~\ref{fig:spec_view}, according to the PSD-based model, the interference caused by the secondary FB-MC system onto the CP-OFDM incumbent is simply obtained by integrating the PSD of the secondary signal on the band $\mathcal{M}_\textsc{I}$ of the incumbent CP-OFDM system. Defining $I_\text{PSD}^{\textsc{S}\rightarrow\textsc{I}}$ the interference injected by the secondary system onto the incumbent one according to the PSD based model, it is expressed as 
\begin{equation}
I_\text{PSD}^{\textsl{S}\rightarrow\textsl{I}} = \int\limits_{\mathcal{M}_\textsl{I}} \Phi_\textsl{S}(f) \dif f,
\end{equation}
where $\Phi_\textsl{S}$ represents the PSD of the secondary signal modulated by secondary system $\textsl{S}$.

Considering that the secondary system modulates i.i.d. symbols with unitary power on each subcarrier, its PSD is defined as 
\begin{equation}
\forall f \in \mathbb{R}, \Phi_\textsl{S}(f) = \sum\limits_{m \in \mathcal{M}_\textsl{S}}\Phi_{m}(f) = \sum\limits_{m \in \mathcal{M}_\textsl{S}}\Phi_{\text{WF},\textsl{S}}(f-m\Delta\text{F}) ,
\end{equation}
where $\Phi_{m,\textsl{S}}(f)$ is the PSD of the $m$th subcarrier of system $\textsl{S}$, and is directly obtained from $\Phi_{\text{WF},\textsl{S}}$, the PSD of one subcarrier of the waveform used by the secondary system, as presented in Fig.~\ref{fig:PSD_waveforms}.
Therefore, it follows that, according to the PSD-based model, 
\begin{equation}
I_\text{PSD}^{\textsl{S}\rightarrow\textsl{I}} = \sum\limits_{m_s \in \mathcal{M}_\textsl{S}}\sum\limits_{m_i \in \mathcal{M}_\textsl{I}} I_\text{PSD}^{\textsl{S}\rightarrow\textsl{I}}(m_s - m_i), 
\end{equation}
with 
\begin{equation}
\forall l \in \mathbb{Z}, I_\text{PSD}^{\textsl{S}\rightarrow\textsl{I}}(l) = \int\limits_{l-\frac{\Delta\text{F}}{2}}^{l+\frac{\Delta\text{F}}{2}}\Phi_{\text{WF},\textsl{S}}(f) \dif f.
\end{equation}
With this notation, $I_\text{PSD}^{\textsl{S}\rightarrow\textsl{I}}(l)$ corresponds to the amount of interference injected by a certain subcarrier $m_s$ of the secondary onto a subcarrier $m_i$ of the incumbent such that $m_s - m_i = l$.

\begin{figure}[t]
	\centering
	\subfloat[]{\includegraphics[trim={1cm 0 1cm 0},clip,width=\linewidth]{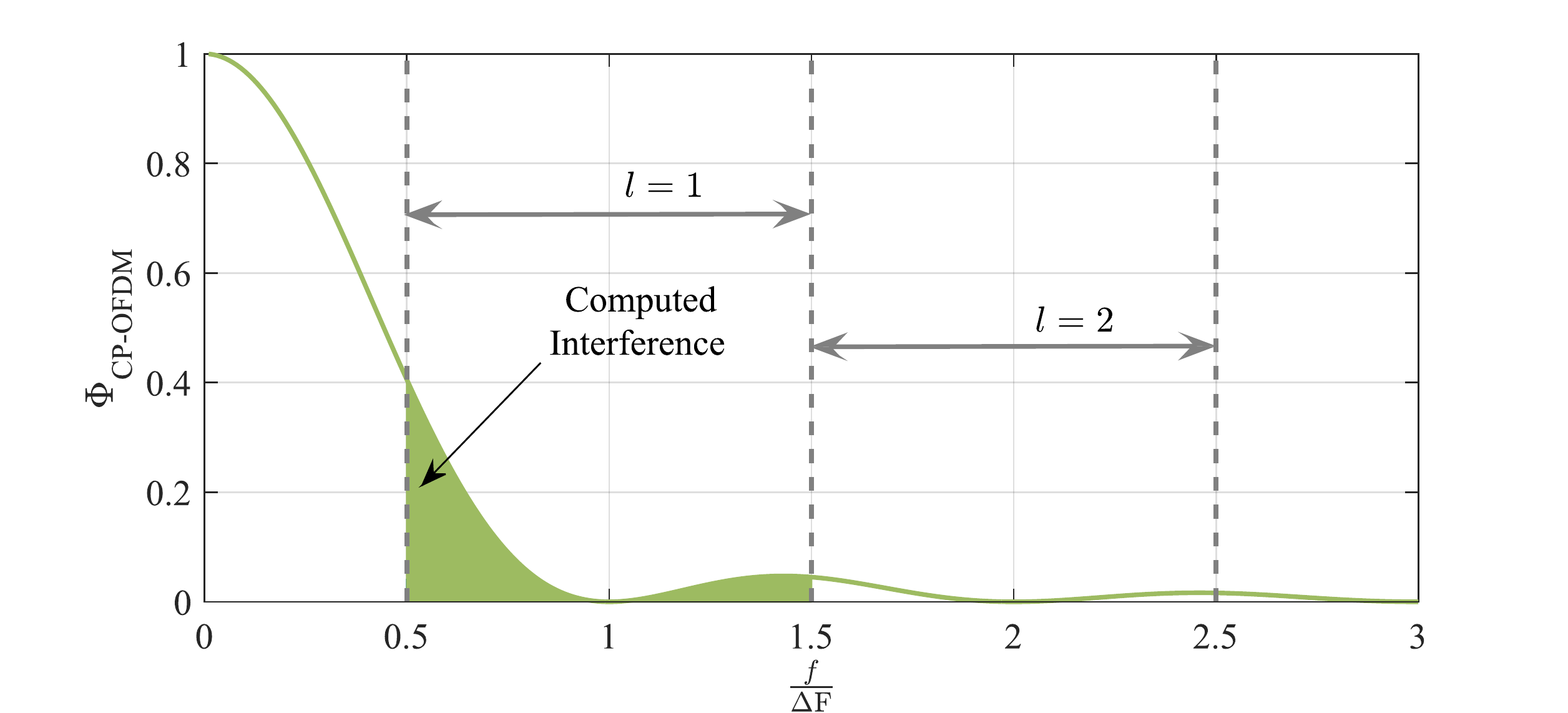}}\\
	\subfloat[]{\includegraphics[trim={1cm 0 1cm 0},clip,width=\linewidth]{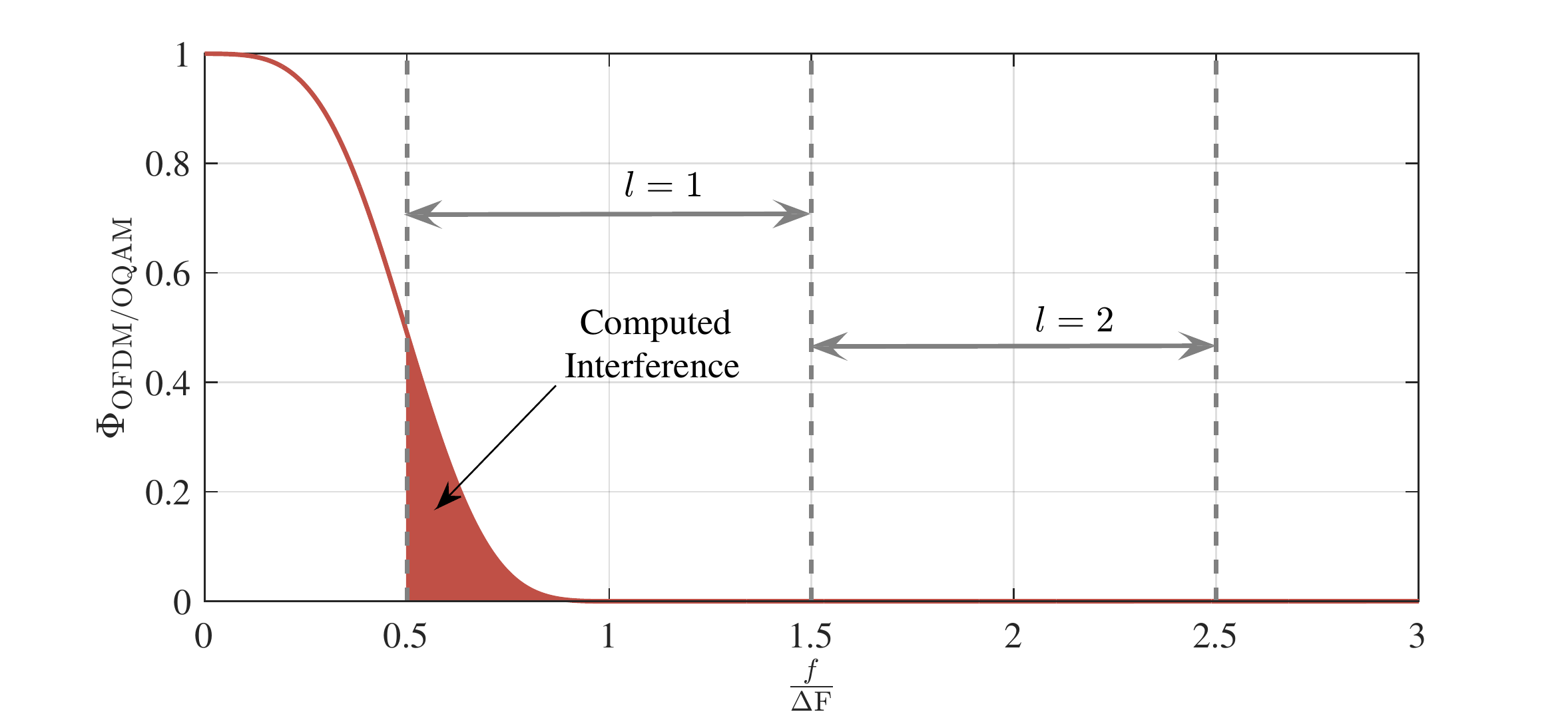}}\\
	\caption{Principle of PSD-based modeling of interference, applied to two waveforms. (a)~:~CP-OFDM, (b)~:~OFDM/OQAM. It applies in the exact same way to other waveforms analyzed in this article.}
	\label{fig:PSD_model_principle}
\end{figure}

Following the symmetric reasoning, the total interference caused by the incumbent system on the secondary is expressed, according to the PSD-based model, as
\begin{equation}
	I_\text{PSD}^{\textsl{I}\rightarrow\textsl{S}} = \sum\limits_{m_s \in \mathcal{M}_\textsl{S}}\sum\limits_{m_i \in \mathcal{M}_\textsl{I}} I_\text{PSD}^{\textsl{I}\rightarrow\textsl{S}}(m_s - m_i),
\end{equation}
and 
\begin{equation}
\forall l \in \mathbb{Z}, I_\text{PSD}^{\textsl{I}\rightarrow\textsl{S}}(l) = \int\limits_{l-\frac{\Delta\text{F}}{2}}^{l+\frac{\Delta\text{F}}{2}}\Phi_{\text{WF},\textsl{I}}(f) \dif f.
\end{equation}

An illustration of how the PSD-based model is used to compute the interference caused by the CP-OFDM and OFDM/OQAM waveforms is given in Fig.~\ref{fig:PSD_model_principle}. We show that the interference caused by the subcarrier of index $0$ onto the subcarrier of index $l$ is computed by integrating the PSD of subcarrier $0$ on the width corresponding to subcarrier $l$. Note that this model applies exactly in the same way to the other FB-MC waveforms we analyze in this study. 

\subsection{Results obtained by each waveform}

\begin{table}

	\caption{Interference tables in dB computed according to the PSD-based model for studied waveforms.}
	\label{tab:PSD_model_results}
	\begin{tabularx}{\linewidth}{|c||m|m|m|m|m|m|}
		\hline
		$l$ & CP-OFDM & FMT & OFDM/\newline OQAM & FBMC-PAM & GFDM & COQAM
	\end{tabularx}		
		\setlength{\tabcolsep}{0pt} 
	\renewcommand{\arraystretch}{0}
	\begin{tabularx}{\linewidth}{|c||M|M|M|M|M|M|}
		\hline
		\hline
		~~$0$~~ & -1.28 & -1.26 & -0.99  & -2.64 & -1.59 & -1.47 \\
		\hline
		$1$ & -12.3 & -10.8 & -12.3  & -13.1 & -10.0 & -10.3 \\
		\hline
		$2$ & -20.1 & -57.2 & -65.4  & -37.2 & -24.1 & -24.7 \\
		\hline
		$3$ & -23.9 & -62.7 & -80.7  & -45.3 & -28.1 & -28.9 \\
		\hline
		$4$ & -26.4 & -65.7 & -89.6  & -50.6 & -30.8 & -31.5 \\
		\hline
		$5$ & -28.0 & -68.0 & -96.1  & -54.6 & -32.8 & -33.4 \\
		\hline
		$6$ & -29.2 & -70.0 & -101. & -58.0 & -34.2 & -35.0 \\
		\hline
		$7$ & -30.3 & -71.0 & -105. & -60.6 & -35.5 & -36.2 \\
		\hline
		$8$ & -31.4 & -72.2 & -109. & -63.0 & -36.7 & -37.4 \\
		\hline
		$9$ & -32.6 & -73.2 & -112. & -65.1 & -37.8 & -38.5 \\
		\hline
	\end{tabularx}
	\centering\includegraphics[width=\linewidth]{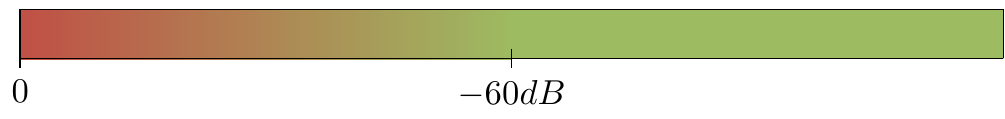}
\end{table}

Having defined the PSD-based model, we present in Table \ref{tab:PSD_model_results} the values of $I_\text{PSD}^{\textsl{S}\rightarrow\textsl{I}}(l)$ obtained for all studied waveforms. As expected, we see that the PSD-based model predicts that waveforms with better spectral localization will inject the lowest amount of interference onto the incumbent CP-OFDM system.

In details, we see clearly that the PSD-based model predicts that OFDM/OQAM, FMT and FBMC-PAM can be used efficiently to coexist with CP-OFDM incumbent systems. On the contrary, GFDM and COQAM systems only reduce the interference onto incumbent systems by $5$ dB compared to CP-OFDM based secondary systems. This is caused by the circular filtering which incurs steep variations in the transmit signal and therefore causes projections on the whole spectrum.

\subsection{Limitations of the PSD-based model}

\begin{figure}
	\subfloat[]{\includegraphics[trim={0.2cm 1cm 0.5cm 1cm},clip,width=0.5\linewidth]{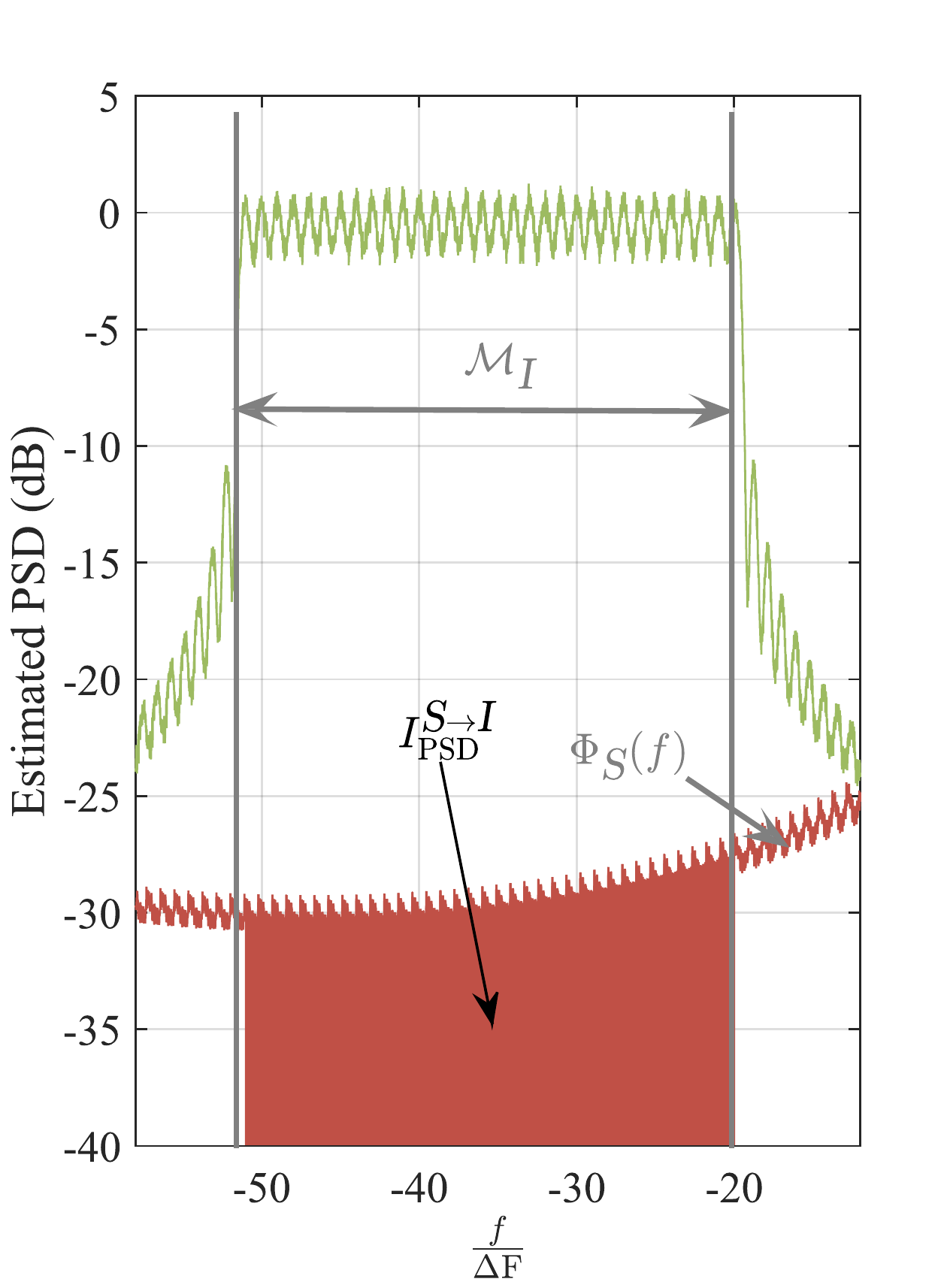}}
	\subfloat[]{\includegraphics[trim={0.2cm 1cm 0.5cm 1cm},clip,width=0.5\linewidth]{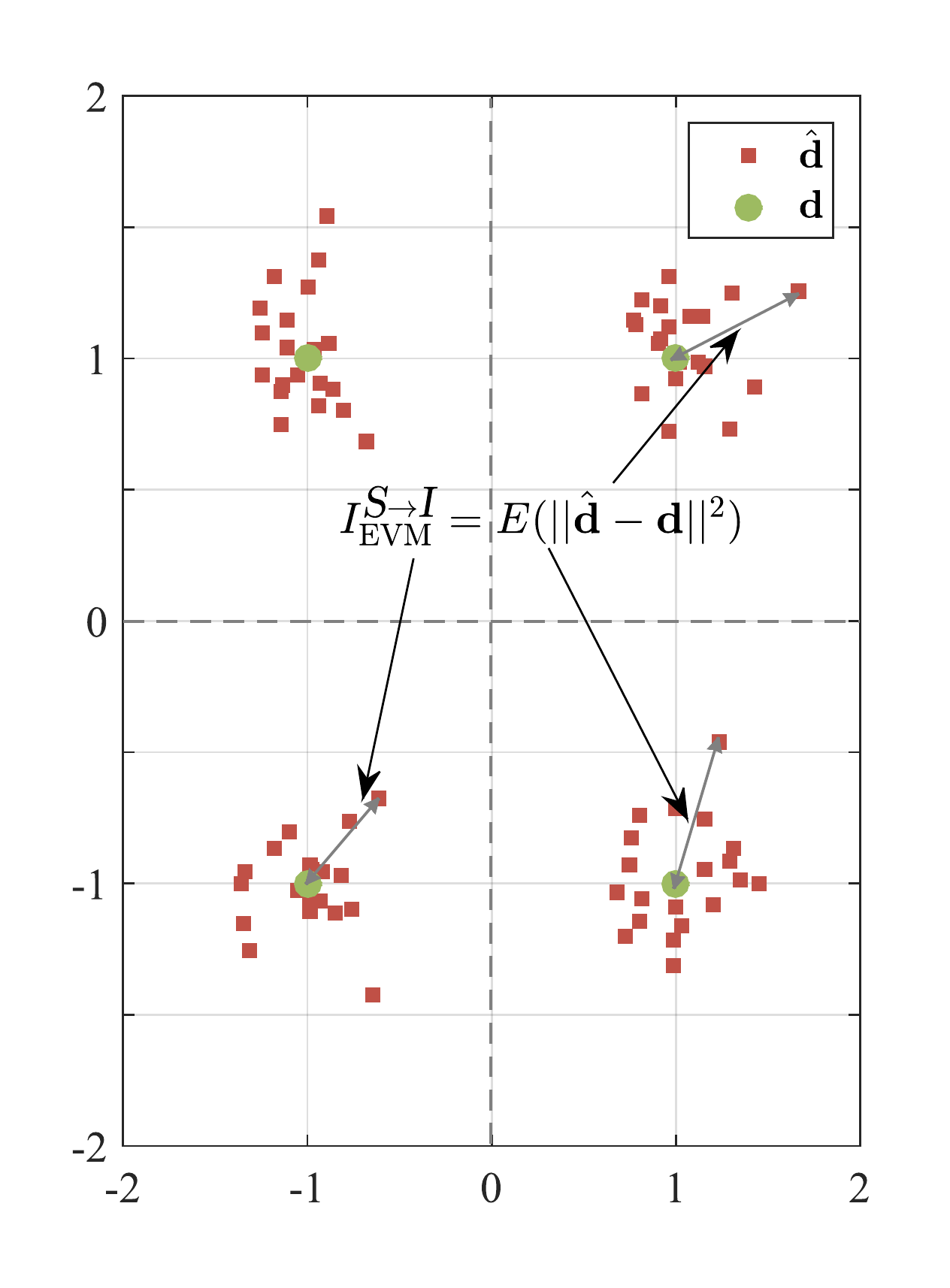}}
	\caption{Comparison between (a) the PSD-based modeling of interference and (b) the EVM-based measure. The PSD-based modeling of interference consists in integrating the PSD of the interfering signal on the band of interest, and therefore does not take into account the receive operations performed at the victim receiver. Interference should rather be measured through EVM at the level of the decoded constellation, just before the symbol decision as shown on (b).}
	\label{fig:difference_approach}
\end{figure}

Though the PSD-based model presented above is widely used in the literature, it suffers some limitations: for example, being based purely on frequency considerations, it is unable to encompass phenomenons related to time synchronism. This was already tackled by Medjahdi \textit{et. al.} in \cite{Medjahdi2010a}. However, the analysis carried out in that article was focused on homogeneous scenarios, in which the coexisting systems were both using the same waveform, either OFDM/OQAM or CP-OFDM. In that scenario, results showed that the main drawback of the PSD-based model lies in the fact that it cannot encompass time asynchronism between coexisting systems. This is a major drawback that does not enable us to analyze the effects of the time misalignment $\delta_\text{t}$ that we introduced in our system model in Section \ref{sec:sys_model}. 

Moreover, it is clear that the interference computed by the PSD-based model $I_\text{PSD}$ is measured \textbf{in the wireless channel}. This means that only the spectral properties of the interfering signal are taken into account. In other words, the PSD-based model computes the interference \textbf{at the input antenna} of the victim receiver, a point marked as $\vcenter{\hbox{\includegraphics[height=1em]{point_A}}}$ in the system model scheme presented in Fig.~\ref{fig:sys_model}. However, the values of interference that really matter are those experienced by the victim receiver after it demodulates the incoming signal, just before the decision, in other words at point $\vcenter{\hbox{\includegraphics[height=1em]{point_B}}}$ in Fig.~\ref{fig:sys_model}. These values of interference are based on measurements of the error vector magnitude (EVM) at the incumbent receiver that suffers from the adjacent transmission of the secondary system, and therefore encompass both the properties of the transmitted secondary signal and the demodulation operations performed by the incumbent receiver. On the opposite, the PSD-based model only accounts for the properties of the transmitted secondary signal and omits the operations performed by the incumbent receiver. An illustration of how the PSD-based and EVM-based measurement of interference differ in their approach is given in Fig.~\ref{fig:difference_approach}. 

It is clear that the EVM-based approach is much more representative of the actual interference experienced by the incumbent receiver. Namely, it is exactly this approach that Medjahdi \textit{et. al.} followed in \cite{Medjahdi2010a} to overcome the limitations of the PSD-based model in their specific scenario where the secondary and incumbent systems use the same waveform. However, in the scenario we study here, getting closed forms of the EVM may be challenging, as it involves much more intricate mathematical derivations than simply integrating the PSD. Therefore, it would be tempting to stick to the PSD based approach in the hope that values given by this model are close enough to the actual EVM values. This is what we propose to (in)validate in the following section by comparing values of interference obtained with the PSD-based model to values obtained through numerical simulations of the EVM.

\section{EVM-based measurement of interference}
\label{sec:interf}
\subsection{Principle of the EVM-based measure of interference}
\label{sec:EVM_principle}
Consider the system model depicted in Fig.~\ref{fig:sys_model}. In the following, parameters indexed as $\cdot_\textsl{I}$ and $\cdot_\textsl{S}$ refer to the incumbent and secondary system respectively. According to \eqref{eq:general_demod} and the expression of the CP-OFDM receive filter given in Section \ref{sec:CPOFDM_param}, at point $\vcenter{\hbox{\includegraphics[height=1em]{point_B}}}$ of Fig.~\ref{fig:sys_model}, the demodulated symbol on the $n_\textsl{I}$th time slot and $m_\textsl{I}$th subcarrier of the incumbent is expressed as
\begin{equation}
\hat{\mathbf{d}}_{m_\textsl{I}}[n_\textsl{I}] = \frac{1}{\sqrt{T_\textsl{I}}}\int\limits_{n_\textsl{I}(T_\textsl{I}+T_\text{CP,\textsl{I}})}^{n_\textsl{I}(T_\textsl{I}+T_\text{CP,\textsl{I}})+T_\textsl{I}}e^{-j2\pi \frac{m_\textsl{I}}{T_\textsl{I}}(t-n_\textsl{I}(T_\textsl{I}+T_\text{CP,\textsl{I}}))}y(t)\dif t
\end{equation}

Recalling the expression of $y(t)$ given in \eqref{eq:sum_in_channel}, after some trivial derivations, we obtain
\begin{equation}
\hat{\mathbf{d}}_{m_\textsl{I}}[n_\textsl{I}] = \mathbf{d}_{m_\textsl{I}}[n_\textsl{I}] + \pmb{\eta}_{m_\textsl{I}}[n_\textsl{I}],
\end{equation}
where $\pmb{\eta}_{m_\textsl{I}}[n_\textsl{I}]$ represent the interference caused by the secondary FB-MC signal and is expressed as 
\begin{align}
\pmb{\eta}_{m_\textsl{I}}[n_\textsl{I}] = \frac{1}{\sqrt{T_\textsl{I}}}\int\limits_{n_\textsl{I}(T_\textsl{I}+T_\text{CP,\textsl{I}})}^{n_\textsl{I}(T_\textsl{I}+T_\text{CP,\textsl{I}})+T_\textsl{I}}&e^{-j2\pi \frac{m_\textsl{I}}{T_\textsl{I}}(t-n_\textsl{I}(T_\textsl{I}+T_\text{CP,\textsl{I}}))}\nonumber\\&\times s_\textsl{S}(t-\delta_{\text{t}})e^{j2\pi\delta_\text{f}(t-\delta_{\text{t}})}\dif t.
\label{eq:eq14}
\end{align}

Then, inserting \eqref{eq:signal_sum_subcarriers} in \eqref{eq:eq14}, and operating the change of variable $t\mapsto t-n_\textsl{I}(T_\textsl{I}+T_\text{CP,\textsl{I}})$, we obtain
\begin{align}
\pmb{\eta}_{m_\textsl{I}}[n_\textsl{I}] = \sum\limits_{m_\textsl{S}\in \mathcal{M}_\textsl{S}}\sum\limits_{n_\textsl{S}\in\mathbb{Z}}&\mathbf{d}_{m_\textsl{S}}[n_\textsl{S}]\frac{1}{\sqrt{T_\textsl{I}}}\int_0^{T_\textsl{I}}\big[f_{n_\textsl{S},\text{T}}(t-\tau(n_\textsl{S},n_\textsl{I}))\nonumber\\&e^{j2\pi[ \nu(m_\textsl{S}, m_\textsl{I}, \delta\text{f}) t+\phi(m_\textsl{S}, m_\textsl{I}, \delta\text{f},\delta\text{t})]}\big]\dif t,
\label{eq:eq15}
\end{align}
with
\begin{align}
\tau(n_\textsl{S},n_\textsl{I}) &= n_\textsl{S}\Delta\text{T}_\textsl{S}- n_\textsl{I}\Delta\text{T}_\textsl{I}+\delta\text{t}\label{eq:tau}\\
\phi(m_\textsl{S}, m_\textsl{I}, \delta\text{f},\delta\text{t}) &= -(m_\textsl{S}\Delta\text{F}_\textsl{S}+\delta\text{f})(n_\textsl{S}\Delta\text{T}_\textsl{S}- n_\textsl{I}\Delta\text{T}_\textsl{I})-\delta\text{f}\delta\text{t}\\
\nu(m_\textsl{S}, m_\textsl{I}, \delta\text{f}) &= m_\textsl{S}\Delta\text{F}_\textsl{S}-m_\textsl{I}\Delta\text{F}_\textsl{I}+\delta\text{f}\label{eq:nu}
\end{align}

Naming $C(m_\textsl{S}, m_\textsl{I}, n_\textsl{S}, n_\textsl{I}, \delta\text{f},\delta\text{t})$ the integral term in \eqref{eq:eq15}, we have
\begin{equation}
\pmb{\eta}_{m_\textsl{I}}[n_\textsl{I}] = \sum\limits_{m_\textsl{S}\in \mathcal{M}_\textsl{S}}\underset{\pmb{\eta}_{m_\textsl{S}\rightarrow m_\textsl{I}}[n_\textsl{I}]}{\underbrace{\sum\limits_{n_\textsl{S}\in\mathbb{Z}}\frac{\mathbf{d}_{m_\textsl{S}}[n_\textsl{S}]}{\sqrt{T_\textsl{I}}}C(m_\textsl{S}, m_\textsl{I}, n_\textsl{S}, n_\textsl{I}, \delta\text{f},\delta\text{t})}}
\end{equation}

Then, the interference injected by the $m_\textsl{S}$th subcarrier of the secondary system onto the $m_\textsl{I}$th subcarrier of the incumbent system during the $n_\textsl{I}$th time-slot is expressed, according to our EVM approach, as 
\begin{align}
\mathbf{I}_\text{EVM}^{\textsl{S}\rightarrow \textsl{I}}(m_\textsl{S},m_\textsl{I},&\delta\text{f},\delta\text{t})[n_\textsl{I}]= E_{\mathbf{d_\textsl{S}}}\{|\pmb{\eta}_{m_\textsl{S}\rightarrow m_\textsl{I}}[n_\textsl{I}]|^2\}\\
&= \frac{\sigma_{\mathbf{d}_{m_\textsl{S}}}^2}{{T_\textsl{I}}}\sum\limits_{n_\textsl{S}\in\mathbb{Z}}|C(m_\textsl{S}, m_\textsl{I}, n_\textsl{S}, n_\textsl{I}, \delta\text{f},\delta\text{t})|^2
\label{eq:eq21}
\end{align}

Finally, in line with the approach in \cite{Medjahdi2010a}, we consider that $\delta\text{t}$ is a random variable that is uniformly distributed in $[0, T_\textsl{i}+T_{\text{CP},\textsl{I}}]$. This is done in order to encompass the lack of synchronization and coordination between the secondary and incumbent systems. Note that we keep a fixed value of $\delta\text{f}$ as this parameter only shifts the secondary transmission in frequency.
The final average value of  measured interference is then obtained as the average EVM for all values of $n_\textbf{I}$ and all possible realisations of $\delta\text{t}$ as 
\begin{equation}
I_\text{EVM}^{\textsl{S}\rightarrow \textsl{I}}(m_\textsl{S},m_\textsl{I},\delta\text{f}) = E_{\delta\text{t}}\left\{\overline{\mathbf{I}_\text{EVM}^{\textsl{S}\rightarrow \textsl{I}}(m_\textsl{S},m_\textsl{I},\delta\text{f},\delta\text{t})}\right\}.
\label{eq:eq22}
\end{equation}

\subsection{Simulation setup and obtained results}
\begin{figure}
	\includegraphics[trim={1cm 1cm 1cm 2cm}, clip, width=\linewidth]{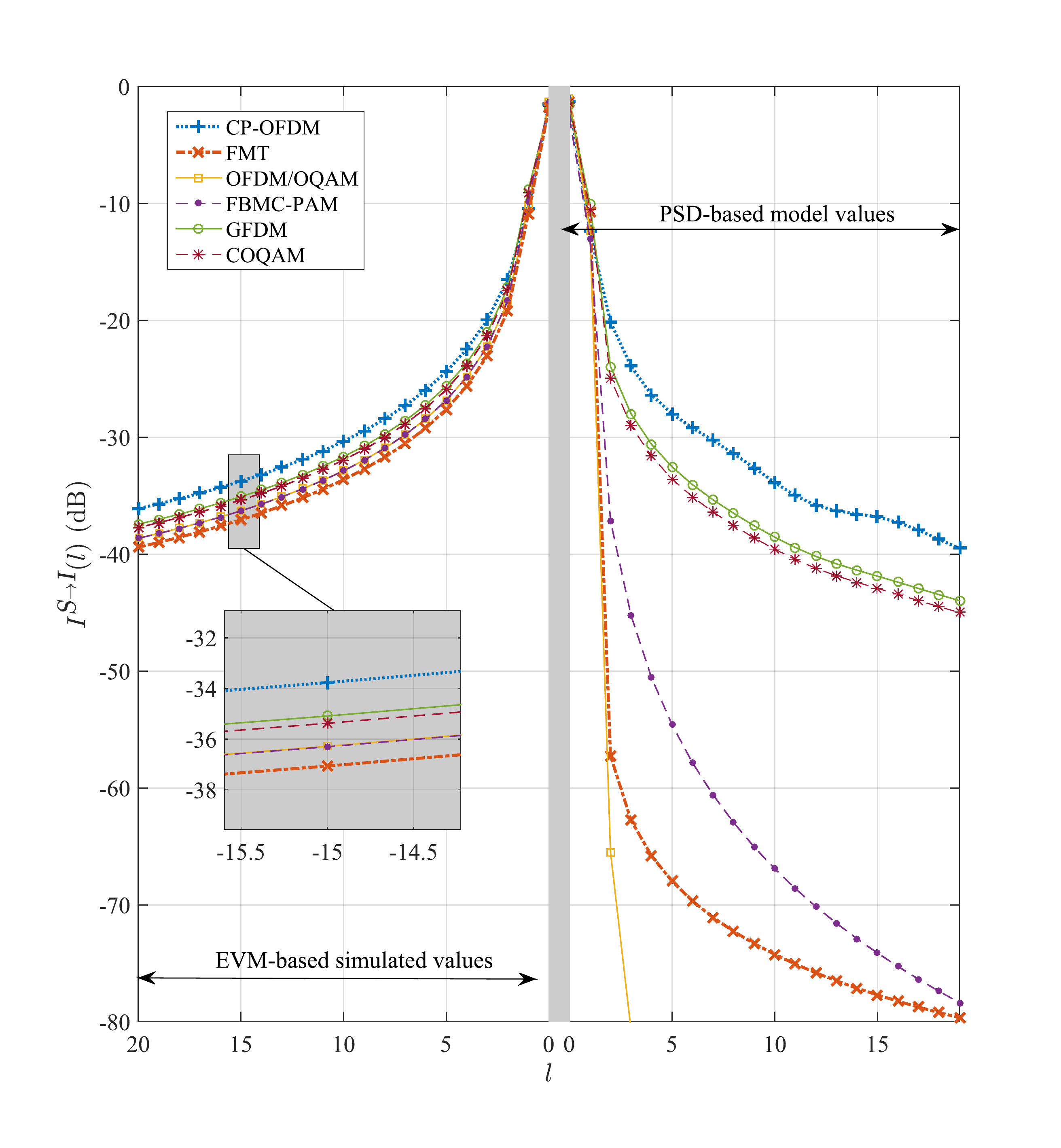}
		\caption{Interference caused onto a CP-OFDM based incumbent system by a secondary system based on multiple waveforms. Values predicted by the PSD-based model (right-hand side) are compared to those obtained through Monte-Carlo simulations of the average EVM at the incumbent CP-OFDM receiver according to \eqref{eq:eq22} (left-hand side). Whatever the waveform used by the secondary system, EVM-based values are dramatically higher than those predicted by the PSD-based model.}
	\label{fig:comp_PSD_EVM}
\end{figure}
Though the EVM-based modeling of interference we presented is much more rigorous than the PSD-based approach, it is also much more convoluted and far less practical. In order to check if it is necessary to pursue the EVM-based modeling of interference, we simulate values obtained according to the model defined in \eqref{eq:eq22}. To do so, we consider the waveforms setups depicted in Table \ref{tab:waveforms_par} and we set $\delta\text{f} = 0$. Thus, as in the PSD-based model, only the value of $m_\textsl{S}-m_\textsl{I} = l$ matters. For each waveform used by the secondary system, we measure the interference experienced on each subcarrier of the incumbent CP-OFDM system according to \eqref{eq:eq22}. 

Obtained results are depicted in Fig.~\ref{fig:comp_PSD_EVM}. Note that, in that figure, we plot both the results obtained with the EVM-based approach on the left side, and those obtained with the PSD-based model on the right side. Undeniably, the actual interference based on EVM measurements after the CP-OFDM demodulator is tremendously higher than what is predicted with the PSD-based model, at the input antenna of the incumbent receiver. The case of OFDM/OQAM based secondary systems gives a glaring example : at a subcarrier distance $l = 2$, the PSD-based model predicts that OFDM/OQAM will inject an interference power of $-65.4$ dB. In fact, simulated EVM values show that it is actually around $-18.4$ dB. This means that the PSD-based model underestimates more than 10~000 times some values of the interference experienced by the CP-OFDM incumbent system. Moreover, we see that the gains achieved by the secondary system if it uses FB-MC waveforms are dramatically reduced. Indeed, the interference caused onto the CP-OFDM incumbent receiver is only reduced by 1 to 3 dB if the secondary system uses a FB-MC waveform instead of CP-OFDM. Interestingly, the order of magnitude of this performance gap is quite in line with the results obtained by some of the experiments we cited in introduction, in particular \cite{EurecomExp, EURECOM+4725}. This brings two conclusions:
\begin{enumerate}
	\item The PSD-based model evaluates the interference to CP-OFDM systems poorly and gives misleading results
	\item Opposed to the common way of thinking, FB-MC waveforms do not drastically facilitate coexistence with incumbent CP-OFDM based legacy systems.
\end{enumerate}

\begin{figure}[t]
	\centering
	\includegraphics[width=\linewidth]{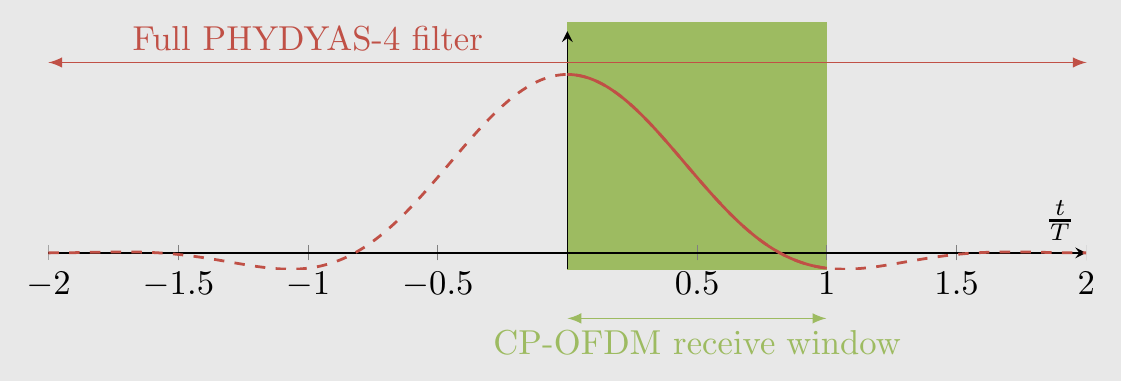}
	\caption{Key phenomenon explaining the difference between the EVM-based measurement of interference and the predictions of the PSD-based model: FB-MC waveforms use longer-than-$T$ filters to achieve high spectral localization, but these are cut by the $T$-long receive window of the incumbent CP-OFDM system. Figure shows the example of an OFDM/OQAM based secondary with PHYDYAS-4 filter, but the same applies for any other FB-MC waveform.}
	\label{fig:physical_explanation}
\end{figure}

\begin{figure*}[b]
	{\footnotesize
	\begin{align}
	\hline\nonumber\\
	|A_{\Pi_{T_\textsl{I}},g}(\tau,\nu)| &=\begin{cases}
	0, &\tau > T_\textsl{I}+\frac{T_g}{2} \text{ or } \tau < -\frac{T_g}{2}.\\
	\left|(\tau+\frac{T_g}{2})\sum\limits_{k \in \mathbb{Z}}G_ke^{j\pi\frac{k}{T_g}(\frac{T_g}{2}-\tau)}\text{sinc}(\pi(\frac{k}{T_g}+\nu)(\frac{T_g}{2}+\tau))\right|, &-\frac{T_g}{2}\leq \tau \leq T_\textsl{I}-\frac{T_g}{2}\\
	\left|(T_\textsl{I}-(-\frac{T_g}{2}+\tau))\sum\limits_{k \in \mathbb{Z}}G_ke^{j\pi\frac{k}{T_g}(-\frac{T_g}{2}-\tau+T_\textsl{I})}\text{sinc}(\pi(\frac{k}{T_g}+\nu)(T_\textsl{I}-(-\frac{T_g}{2}+\tau)))\right|, &\frac{T_g}{2} \leq \tau \leq T_\textsl{I}+\frac{T_g}{2}\\
	\left|T_\textsl{I}\sum\limits_{k \in \mathbb{Z}}G_ke^{j\pi\frac{k}{T_g}(T_\textsl{I}-2\tau)}\text{sinc}(\pi(\frac{k}{T_g}+\nu)T_\textsl{I})\right|, & T_\textsl{I}-\frac{T_g}{2} \leq \tau \leq \frac{T_g}{2}
	\end{cases}\label{eq:A_pi_t_g}
	\end{align}}
\end{figure*}

So, why is the PSD-based model so inaccurate? The main reason is given in Fig.~\ref{fig:physical_explanation} : FB-MC systems rely on prototype filters that are longer than the time-symbol to achieve acute frequency localization. However, the CP-OFDM receiver filter is a rectangular window of length equal to the time-symbol. Therefore, it truncates the well-shaped prototype filter used to transmit the FB-MC secondary signal in parts that, on their own, are not particularly well frequency localized. 
As a result of this operation, the advantageous PSD properties of the FB-MC signals are lost by the CP-OFDM incumbent receiver. Note that this is a general result that can be extended to virtually any signal passing through a CP-OFDM receiver. Indeed, the latter acts exactly as a poor spectrum analyzer which outputs high frequency ripples even when fed a well localized signal.

\section{Analytical aspects}
\label{sec:analysis}

At this point of our discussion, the main ideas of this article have been exposed. To finish convincing the reader of the validity of our claims, we derive in this section the mathematical analysis of the inter-system interference according to the mathematical model laid out in Section \ref{sec:EVM_principle}. 
Let us fix $m_\textsl{I}$, $m_\textsl{S}$, $\delta\text{f}$ and $\delta\text{t}$.
In \eqref{eq:eq21}, $I_\text{EVM}^{\textsl{S}\rightarrow \textsl{I}}(m_\textsl{S},m_\textsl{I},\delta\text{f},\delta\text{t}) $ is written as the sum of specific values of  the magnitude of the cross-ambiguity function between the transmit filter of the secondary and a rectangular window. 

In the case where the secondary system is based on linear convolution FB-MC, we reminded in Section \ref{sec:linear_FBMC} that the transmit filter used to modulate each subsequent symbol is equal to the used prototype filter $g$. 
Therefore, we can show that $|C(m_\textsl{S}, m_\textsl{I}, n_\textsl{S}, n_\textsl{I}, \delta\text{f},\delta\text{t})|$ can be expressed as $\left|A_{\Pi_{T_\textsl{I}},g}(\tau(n_\textsl{S}, n_\textsl{I}),\nu(m_\textsl{S},  m_\textsl{I}, \delta\text{f}))\right|$, the cross-ambiguity function between a rectangular window of length $T_\textsl{I}$ and the prototype filter $g$, with $\tau$ and $\nu$ following the expressions of \eqref{eq:tau},\eqref{eq:nu}. The expression of $|A_{\Pi_T,g}|$ is given in \eqref{eq:A_pi_t_g} where the terms $G_k$ represent the Fourier coefficients of the prototype filter $g$ used by the secondary system. Developments leading to this expression are detailed in Appendix. 
Therefore, \eqref{eq:eq21} is rewritten as
{\small
\begin{equation}
\mathbf{I}_\text{EVM}^{\textsl{S}\rightarrow \textsl{I}}(m_\textsl{S},m_\textsl{I},\delta\text{f},\delta\text{t})[n_\textsl{I}]
= \frac{\sigma_{\mathbf{d}_{m_\textsl{S}}}^2}{{T_\textsl{I}}}\sum\limits_{n_\textsl{S}\in\mathbb{Z}}|A_{\Pi_{T_\textsl{I}},g}(\tau(n_\textsl{S},n_\textsl{I}),\nu(m_\textsl{S},m_\textsl{I},\delta_\text{f}))|^2
\label{eq:eq24}
\end{equation}}%

Putting \eqref{eq:A_pi_t_g} in this last expression, we see clearly that the power of interference caused by the secondary FB-MC waveform onto the incumbent CP-OFDM system slowly decays in frequency, following a weighted sum of $\text{sinc}$ functions. This last expression can be used to rate the exact interference seen on the $n_\textsl{I}$th time slot of the incumbent system. This gives us a closed-form expression of the interference seen on each symbol of the incumbent CP-OFDM system, for a particular value of time misalignment $\delta_\text{t}$ between the two systems.

However, obtaining closed-form expressions of the average interference as expressed in \eqref{eq:eq22} is impractical, because it involves taking the average value of the vector $\mathbf{I}_\text{EVM}^{\textsl{S}\rightarrow \textsl{I}}(m_\textsl{S},m_\textsl{I},\delta\text{f},\delta\text{t})$, and then take the expected value of the result with respect to random variable $\delta_\text{t}$. Moreover, the results presented in Fig.~\ref{fig:comp_PSD_EVM} reduce the appeal for closed-form expressions. The main point we want to show in this analysis is that the CP-OFDM reception causes high interference to the incumbent system , which is explained by the appearance of the sinc function in \eqref{eq:A_pi_t_g}. 

Nevertheless, based on our previous works in \cite{BodinierICC2016, Bodinier2016GC}, we can assert that the value of interference is in most cases only slightly dependent on the values of $n_\textsl{I}$ and $\delta\text{t}$. Therefore, in order to offer tractable equations, it is possible to adopt the following approximation:
\begin{equation}
I_\text{EVM}^{\textsl{S}\rightarrow \textsl{I}}(m_\textsl{S},m_\textsl{I},\delta\text{f}) \approx \mathbf{I}_\text{EVM}^{\textsl{S}\rightarrow \textsl{I}}(m_\textsl{S},m_\textsl{I},\delta\text{f},\delta\text{t}=0)[0]
\end{equation}
With this approximated form, the expression of \eqref{eq:eq22} simplifies into
\begin{align}
I_\text{EVM}^{\textsl{S}\rightarrow \textsl{I}}(m_\textsl{S},m_\textsl{I},\delta\text{f}) \approx \sigma_{\mathbf{d}_{m_\textsl{S}}}^2 {T_\textsl{I}}\sum\limits_{n=0}^{\frac{T_g}{\Delta\text{T}_\textsl{S}}} \sum\limits_{k \in \mathbb{Z}} & G_ke^{j\pi\frac{k}{T_\text{g}}(T_\textsl{I}-2(n\Delta\text{T}_\textsl{S}))}\nonumber\\
&\times \text{sinc}(\pi(\frac{k}{T_\text{g}}+\nu)T_\textsl{I})
\label{eq:eq25}
\end{align}

Note that in the case where the secondary system  uses circular convolution filter banks, the expression of \eqref{eq:A_pi_t_g} is not directly applicable. Indeed, because of the CP addition, as shown in Fig.~\ref{fig:lineartocirc}, subsequent symbols pass through different filters that have slightly different Fourier coefficients. However, the main principle stays unchanged and $\eqref{eq:eq25}$ can be used as an approximation as well in the case of circular convolution FB-MC waveforms.

\begin{figure}[t]
\begin{picture}(220,220)
	\put(0,10){\includegraphics[width=\linewidth]{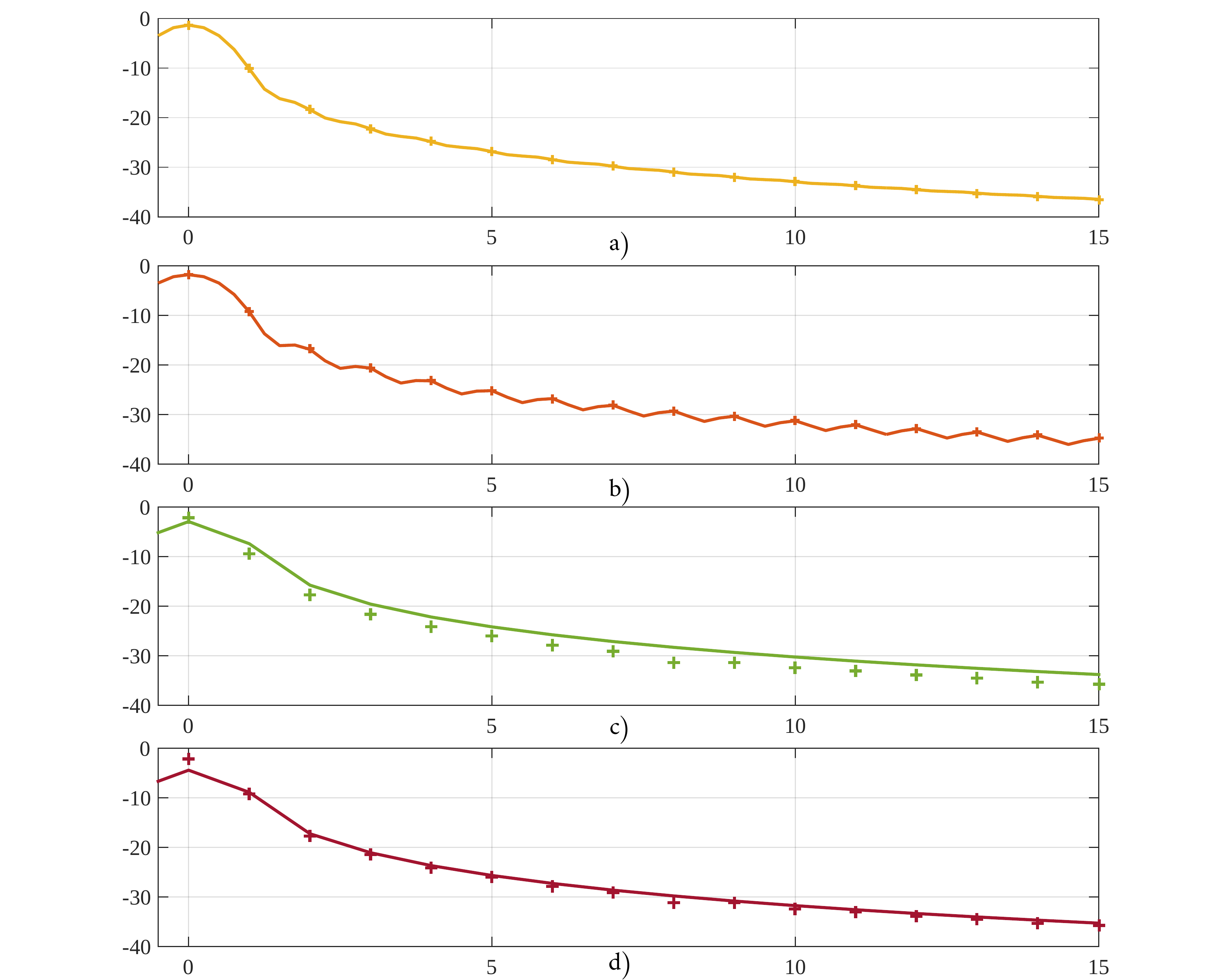}}
	\put(5,70){\rotatebox{90}{\footnotesize$I_\text{EVM}^{\textsl{S}\rightarrow \textsl{I}}(m_\textsl{S},m_\textsl{I},\delta\text{f})\ (dB)$}}
	\put(100, 0){\footnotesize$\nu(m_\textsl{S},m_\textsl{I},\delta\text{f})$}
\end{picture}
\caption{Comparison between simulated values of average interference based on EVM (crosses) and analytical approximation of \eqref{eq:eq25} (solid line) for a) OFDM/OQAM, b) FMT, c) GFDM and d) COQAM.}
\label{fig:comp_theo}
\end{figure}

In Fig. \ref{fig:comp_theo}, we compare the approximation of $I_\text{EVM}^{\textsl{S}\rightarrow \textsl{I}}(m_\textsl{S},m_\textsl{I},\delta\text{f})$ given in \eqref{eq:eq25} with results obtained through Monte-Carlo simulations. We see that the proposed approximation matches perfectly the simulated values for linear filter banks but only approximates the interference values in the case of circular filter banks, as predicted. Nevertheless, the obtained approximation is satisfying and shows well, for each waveform, that the interference to the CP-OFDM incumbent slowly decays, at approximately the same rate no matter what waveform is used by the secondary system. Note that we did not represent the curve for FBMC-PAM as it is clear from Fig.~\ref{fig:comp_PSD_EVM} that it interferes exactly as much as OFDM/OQAM.

\begin{figure}[t]
	\includegraphics[trim={2cm 0cm 2cm 0.2cm}, clip, width=\linewidth]{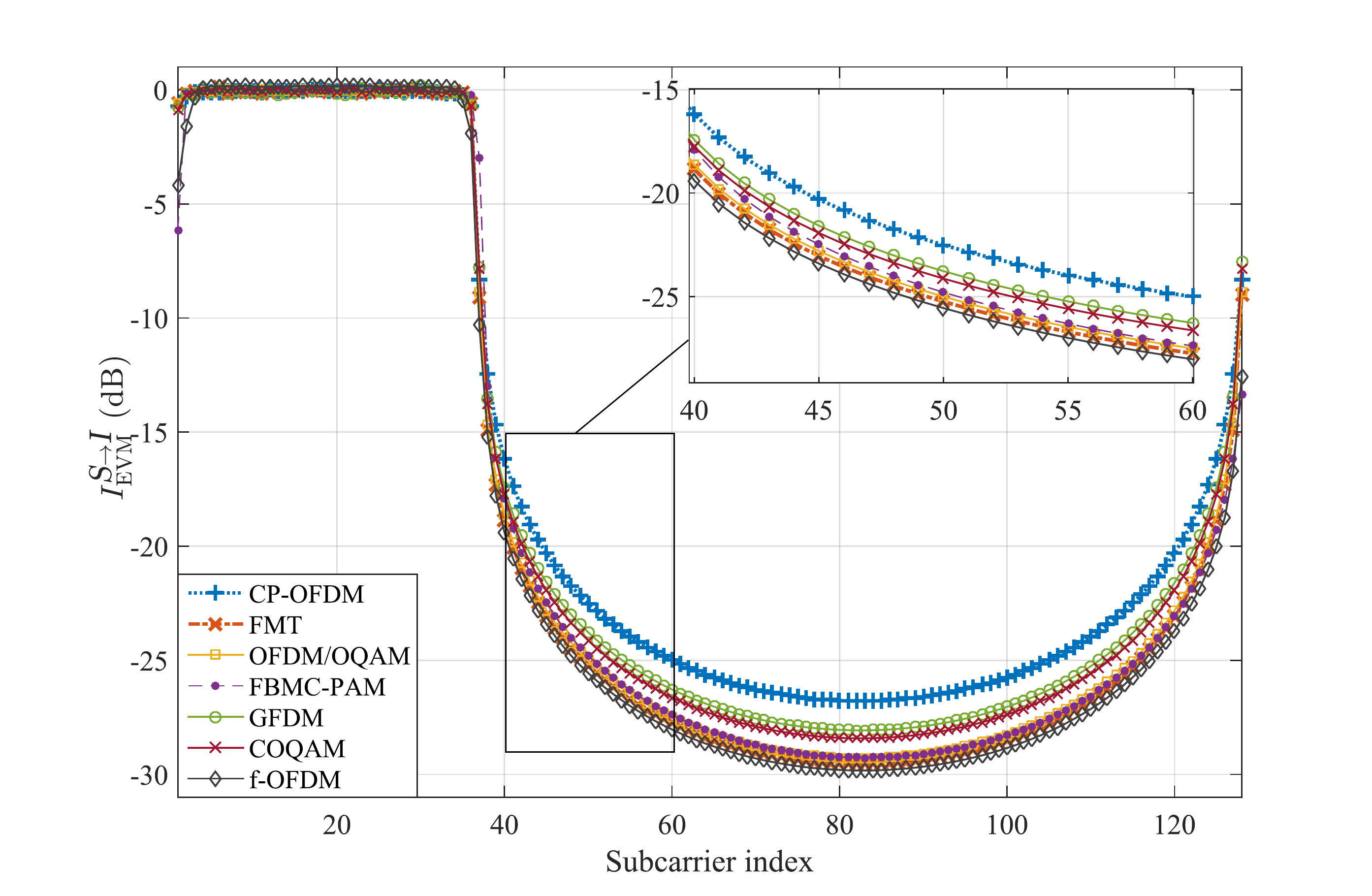}
		\caption{EVM on each subcarrier of the CP-OFDM incumbent receiver when a secondary user is transmitting on subcarriers $0$ to $35$ with different waveforms. FB-MC waveforms follow the parameters listed in Table \ref{tab:waveforms_par} and f-OFDM those in \cite{Abdoli2015}. f-OFDM does not bring any significant advantage compared to analyzed FB-MC waveforms.}
	\label{fig:fOFDM}
\end{figure}

At this point of our discussion, we have made clear that all FB-MC waveforms that we have studied in this article fail to properly coexist with CP-OFDM incumbent systems, and shed light on the core analytical aspects that are responsible for this. Mainly, we have explained that the advantageous spectral properties of FB-MC waveforms are drastically reduced when the long prototype filter is truncated by the rectangular window of the CP-OFDM receiver. 
However, a limitation of our work is that it is focused solely on FB-MC waveforms that perform a filtering per subcarrier. This was done on purpose to maintain a common mathematical model that would allow us to obtain tractable expressions. Therefore, one may wonder if the limitations shown in this article cannot be overcome if the secondary uses a subband filtered waveform such as UF-OFDM \cite{Wild2014} or f-OFDM \cite{Abdoli2015}. Indeed, these waveforms use wider, hence shorter, filters. However, even with such a waveform, the CP-OFDM reception truncates the incoming signal with the same rectangular window and the same problem appears as for the FB-MC waveforms we have analyzed in that article. To make it clear, we present in Fig.~\ref{fig:fOFDM} the EVM on each subcarrier of the CP-OFDM incumbent receiver when the secondary user is active on subcarriers $0$ to $35$. We compare all FB-MC waveforms analyzed in this article with a f-OFDM system parameterized as in \cite{Abdoli2015}. We show clearly that f-OFDM and FB-MC waveforms interfere to a similar extent on the CP-OFDM incumbent receiver. This shows once more that the limiting factor in the scenario we study is the CP-OFDM receiver, that truncates the incoming signal and outputs a poor spectral representation of it.

\section{Conclusion}
In this paper, we investigated the coexistence on the same
band of users utilizing FB-MC waveforms and legacy CP-OFDM incumbent devices. We first reminded
that in these scenarios, interference between users with different
physical layers is usually measured according to the
PSD of the interfering signal. This PSD-based model, though practical, only encompasses the effects of interference in the channel, at the input antenna of the receiver that suffers from interference. 

However, interference should actually be measured based on EVM after the demodulation operations, before the constellation decoding. We therefore compared values of interference measured at this stage with those predicted by the PSD-based model.
We showed that there is a tremendous gap between the PSD-based and EVM-based model, and went on to explain that this is due to the fact that the CP-OFDM receiver truncates the incoming interfering signal, which destroys the advantageous PSD properties of FB-MC signals. We further validated these results through mathematical analysis.
The obtained results show that secondary users based on
FB-MC waveforms interfere on CP-OFDM incumbent users in a
similar extent as CP-OFDM based secondary users.
Though the mathematical and analytical considerations behind
this work are not particularly complex, they had been
totally omitted in a vast majority of the works available in the
literature. 

Finally, the study carried out in this article shows that the
in-band coexistence of new 5G communications with legacy
systems will only marginally be improved through the use
of FB-MC waveforms. This improvement would be even more limited if signal impairments caused by the non-linear high power amplifier were taken into account. Other technologies that are currently being studied for 5G, such as the use of Massive MIMO, are
therefore likely to be much more efficient at protecting incumbent legacy
CP-OFDM users.

\label{sec:ccl}


%

\appendix
Consider a rectangular window of length $T$, noted $\Pi_T$ and a filter $g$ of length $T_g$. We assume that $g$ respects the following properties: it is equal to $0$ anywhere except for $t \in [a,\ a+T_g]$ and is Lebesgue integrable on its non null area. Note that this is not a strong assumption and that filters commonly used in the literature respect it. We also add the condition that $T_g > T$. Under these assumptions, it is possible to define $g$ as a truncation of a periodic signal $g_\infty(t)$ defined as
\begin{equation}
g_\infty(t) = g(\frac{t-a}{T_g}), \forall t \in \mathbb{R}.
\end{equation}

With this definition, $g_\infty(t)$ is a $T_g$-periodic signal which is Lebesgue integral on one period. Therefore, it can be decomposed in Fourier series as
\begin{equation}
g_\infty(t) = \sum_{k \in \mathbb{Z}}G_k e^{j2\pi k\frac{t}{T_g}}, t \in \mathbb{R}
\end{equation}
where $G_k$ are the Fourier coefficients of $\tilde{g}$.
Therefore, the cross-ambiguity function of $g$ and a rectangular window of length $T$ is equal to 
\begin{align}
A_{\Pi_T,g}(\tau,\nu) &= \int\limits_{0}^{T}g(t-\tau)e^{j2\pi t\nu}dt,
\\
&= \sum_{k \in \mathbb{Z}}G_ke^{-j2\pi k\frac{\tau}{T_g}}\int\limits_{0}^{T}e^{j2\pi t(\frac{k}{T_g}+\nu)}dt
\end{align}
which simplifies into different expressions according to the values of $\tau$.\\

\subsubsection{$T_g\geq T$ case}
Here, we consider the case $T_g\geq T$. 

\textbf{Case I:} $\tau > T-a$ or $\tau < -(a+T_g)$.

In that case, $g(t-\tau)$ does not overlap the rectangular window and 
\begin{equation}
A_{\Pi_T,g}(\tau,\nu) = 0
\label{eq:caseI}
\end{equation}

\textbf{Case II:} $-(a+T_g)\leq \tau \leq T-(a+T_g)$

In that case, only a small part of the filter overlaps the beginning of the rectangular window and 
\begin{equation}
A_{\Pi_T,g}(\tau, \nu) =\sum_{k \in \mathbb{Z}}G_ke^{-j2\pi k\frac{\tau}{T_g}}\int\limits_{0}^{a+\tau+T_g}e^{j2\pi t(\frac{k}{T_g}+\nu)}dt
\label{eq:caseII}
\end{equation}

\textbf{Case III:} $-a \leq \tau \leq T-a$

In that case, only a small part of the filter overlaps the end of the rectangular window and 
\begin{equation}
A_{\Pi_T,g}(\tau, \nu) =\sum_{k \in \mathbb{Z}}G_ke^{-j2\pi k\frac{\tau}{T_g}}\int\limits_{a+\tau}^{T}e^{j2\pi t(\frac{k}{T_g}+\nu)}dt
\label{eq:caseIII}
\end{equation}

\textbf{Case IV:} $T-(a+T_g) \leq \tau \leq -a$

In that case, the filter $g$ overlaps with the whole rectangular window and 
\begin{equation}
A_{\Pi_T,g}(\tau, \nu) =\sum_{k \in \mathbb{Z}}G_ke^{-j2\pi k\frac{\tau}{T_g}}\int\limits_{0}^{T}e^{j2\pi t(\frac{k}{T_g}+\nu)}dt
\label{eq:caseIV}
\end{equation}

In order to give a simplified expression of $A_{\Pi_T,g}(\tau,\nu)$ in all the listed cases, we derive in the following lines the generic expression of $I_b^c = \int\limits_{b}^{c}e^{j2\pi t(\frac{k}{T_g}+\nu)}$.
\begin{align}
I_b^c &=\frac{e^{j2\pi c(\frac{k}{T_g}+\nu)}-e^{j2\pi b(\frac{k}{T_g}+\nu)}}{j2\pi(\frac{k}{T_g}+\nu)}\\
&= (c-b)e^{j\pi(\frac{k}{T_g}+\nu)(b+c)}\text{sinc}\left(\pi(\frac{k}{T_g}+\nu)(c-b)\right)
\label{eq:Ibc}
\end{align}

\begin{figure*}[t]
	{\footnotesize
	\begin{align}
	A_{\Pi_T,g}(\tau,\nu) &=\begin{cases}
	0, &\tau > T-a \text{ or } \tau < -(a+T_g).\\
	(a+\tau+T_g)e^{j\pi\nu(a+\tau+T_g)}\sum\limits_{k \in \mathbb{Z}}G_ke^{j\pi\frac{k}{T_g}(a-\tau+T_g)}\text{sinc}(\pi(\frac{k}{T_g}+\nu)(a+\tau+T_g)), &-(a+T_g)\leq \tau \leq T-(a+T_g) \\
	(T-(a+\tau))e^{j\pi\nu(a+\tau+T)}\sum\limits_{k \in \mathbb{Z}}G_ke^{j\pi\frac{k}{T_g}(a-\tau+T)}\text{sinc}(\pi(\frac{k}{T_g}+\nu)(T-(a+\tau))), &-a \leq \tau \leq T-a\\
	Te^{j\pi\nu T}\sum\limits_{k \in \mathbb{Z}}G_ke^{j\pi\frac{k}{T_g}(T-2\tau)}\text{sinc}(\pi(\frac{k}{T_g}+\nu)T), &T-(a+T_g) \leq \tau \leq -a
	\end{cases}\label{eq:CrossAmbTsmaller}\\
	\hline\nonumber
	\end{align}}
\vspace{-1cm}
\end{figure*}

Putting $(\ref{eq:Ibc})$ into $(\ref{eq:caseI})-(\ref{eq:caseIV})$, we obtain the expression of (\ref{eq:CrossAmbTsmaller}) for $A_{\Pi_T,g}(\tau,\nu)$.

%
\ifCLASSOPTIONcaptionsoff
  \newpage
\fi

\bibliographystyle{../styles/IEEEtranBST/IEEEtran}
{\fontsize{10pt}{12pt}
\bibliography{../styles/IEEEtranBST/IEEEabrv,library}}

\end{document}